\documentclass[%
 aip,
 amsmath,amssymb,
reprint,
]{revtex4-2}

\usepackage{graphicx}
\usepackage{dcolumn}
\usepackage{bm}

\usepackage[utf8]{inputenc}
\usepackage[T1]{fontenc}
\usepackage{mathptmx}
\usepackage{etoolbox}
\usepackage{hyperref}
\usepackage{lipsum}
\usepackage[caption=false]{subfig}
\usepackage{siunitx}
\usepackage{mathtools}
\usepackage{booktabs}
\usepackage{placeins}
\usepackage{xcolor}

\newcommand{\correct}[1]{{#1}}

\hypersetup{%
    pdfpagemode={UseOutlines},
    bookmarksopen,
    pdfstartview={FitH},
    colorlinks,
    linkcolor={black},
    citecolor={black},
    urlcolor={blue}
  }

\makeatletter
\def\@email#1#2{%
 \endgroup
 \patchcmd{\titleblock@produce}
  {\frontmatter@RRAPformat}
  {\frontmatter@RRAPformat{\produce@RRAP{*#1\href{mailto:#2}{#2}}}\frontmatter@RRAPformat}
  {}{}
}%
\makeatother
\begin{document}

\preprint{AIP/123-QED}

\title[Visibility analysis of boundary layer transition]{Visibility analysis of boundary layer transition}
\author{D. Perrone}
\email{davide.perrone@polito.it}
\affiliation{Department of Mechanical and Aerospace Engineering, Politecnico di Torino, 10129 Turin, Italy}
\author{L. Ridolfi}%
\affiliation{Department of Environmental, Land and Infrastructure Engineering, Politecnico di Torino, 10129 Turin, Italy}%
\author{S. Scarsoglio}
\affiliation{Department of Mechanical and Aerospace Engineering, Politecnico di Torino, 10129 Turin, Italy}%

\date{\today}

\begin{abstract}
We study the transition to turbulence in a flat plate boundary layer by means of visibility analysis of velocity time-series extracted across the flow domain. By taking into account the mutual visibility of sampled values, visibility graphs are constructed from each time series. The latter are thus transformed into a geometrical object, whose main features can be explored using measures typical of network science that provide a reduced order representation of the underlying flow properties. Using these metrics, we observe the evolution of the flow from laminarity to turbulence and the effects exerted by the free-stream turbulence. Differently from other methods requiring an extensive amount of spatio-temporal data (\textit{e.g.}, full velocity field) or a set of parameters and thresholds arbitrarily chosen by the user, the present network-based approach is able to identify the onset markers for transition by means of the streamwise velocity time-series alone.
\end{abstract}

\maketitle

\section{Introduction}
\label{sec:intro}
Turbulence arises from laminar fluid motion following the growth of small perturbations of the velocity profile. The interest in the study of transition is well motivated by the physical and industrial implications associated with the appearance of turbulence, which are preeminently linked to increased momentum transfer and drag. Important efforts in this sense are directed towards understanding, predicting and ultimately controlling transition.

In this work, we focus our attention onto transition in the boundary layer of a flat plate with zero pressure gradient. In this setup, transition usually follows two routes. In one case, the laminar profile first \correct{develops} Tollmien-Schlichting waves, whose secondary instabilities then lead to the breakdown to turbulence. In contrast to this \textit{orderly} pathway, the second case occurs when turbulence in the free stream exerts a vertical forcing on the boundary layer and triggers the transition. This route is termed \textit{bypass} transition and will be the focus of this work. Bypass transition takes place as follows: initially, the perturbations in the free stream generate large scale perturbations of the streamwise velocity, \textit{i.e.} streaks, inside the laminar boundary layer. Indeed, while outside the boundary layer the turbulence spectrum has a broadband nature, inside the boundary layer only low frequency perturbations appear, due to shear-sheltering. The secondary instability of streaks generates turbulent spots that are advected by the mean flow and grow in size, until their growth and merging \correct{results} in the full onset of turbulence \citep{zaki2013ftc}.

While for the orderly transition the onset of turbulence and its spatial location can be somewhat predicted and a definition of a critical Reynolds number is usually accepted, in the case of bypass transition the problem is complicated by the chaotic nature of the forcing introduced by the free stream turbulence, so that the onset of turbulence also depends on the turbulent intensity \cite{vermeersch2010,bhushan2018}.
\correct{In general, the onset of turbulence is influenced by different key factors, such as the system geometry, the surface roughness, the external flow condition (most notably the pressure gradient and the already mentioned freestream turbulence).}
Simplified criteria have been proposed, relying on determining a critical ratio between the production term inside streamwise velocity streaks and viscous dissipation. A closely related problem to the prediction of transition is that of the determination of the spatiotemporal location of the turbulent-non turbulent interface (TNTI). The identification of the TNTI in its simplest forms relies on the observation that certain flow field quantities usually assume different range of values in turbulent and laminar regimes \cite{sreenivasan1986,chauhan2014,borrell2016}. 
As such, providing an indicator function and a threshold value should, in theory, suffice to accurately discriminate between laminar and turbulent regions of the domain. Vorticity is a characterizing feature of turbulence, but its use as an indicator function is problematic due to the presence of free-stream turbulence or laminar regions where vorticity is present nonetheless, such as the streaks in the turbulent boundary layer.
Other choices for the indicator function rely on the presence of velocity fluctuations to act as a discriminant between turbulent and laminar flow. A common choice in flat plate boundary layers is the sum of the wall-normal and spanwise velocity fluctuations $\lvert v' \rvert + \lvert w' \rvert$. Still, the fact that the intensity of the velocity fluctuations is strongly dependent on the distance from the wall complicates the choice of an appropriate threshold. Methods such as Otsu's\cite{otsu1979} can be used to determine an appropriate threshold at different wall-normal heights, but the choice can be prone to error and an extensive knowledge of the flow field is necessary \cite{nolan2013}.

More recently, and fueled by the increasing availability of high-resolution flow simulations and experiments, several data-driven approaches have been proposed, with the main goal of discerning the various states present in a transitional boundary layer and the key processes that govern the shift of the flow field between these states \cite{foroozan2021,wu2019}. 
Machine-learning techniques are a broad category of data analysis tools commonly employed to analyze and classify large datasets and discover underlying relations, whose application to fluid mechanics is gaining traction. Some of the most promising applications in the field of transitional flows regard the classification of flow states using unsupervised deep learning approaches. The main merit of unsupervised approaches lies in eliminating the need for \textit{a priori} defined classifiers and/or threshold values, which are instead obtained by the learning algorithm itself.
On the other hand, these approaches need extensive training on large datasets, thus requiring the knowledge of the entire velocity field and possibly also of reduction techniques to feed the unsupervised classifiers with treatable data in which the underlying relations are more easily discovered. Moreover, the very large number of parameters usually contained in a machine generated model makes somewhat difficult to obtain a clear physical interpretation of the features of classifiers, which would be useful to drawn more general conclusions about the nature of the studied flow.

An alternative to data-driven approaches comes from dimensionality reduction techniques that, when applied to turbulent flows, are able to extract key information from an otherwise extremely complex flow field. Along with more established methods, such as proper orthogonal decomposition, the application of techniques derived from graph theory has seen a wealth of promising applications to fluid dynamics recently \cite{iacobello2021,taira2022}. Graphs, or networks, are mathematical objects composed of a set of nodes and a set of the interactions entertained by the nodes; they are suited to represent large, and complex, dynamical systems, of which they can capture the essential behavior \cite{boccaletti2006}. 
Network-based methods have been applied to fluid flows to investigate the correlation between the velocity at different points in space \cite{scarsoglio2016complex} or at different Lagrangian trajectories\cite{schlueter2017coherent,vieweg}, vortical interactions\cite{taira2016jfm,yeh2021}, the proximity of particle trajectories\cite{rypina2017npg,banisch} and their transition probability between subsets of the domain\cite{sergiac15,perrone20,perrone21}.

Time-series analysis in the context of network-based methods is particularly relevant, as the structure of time-series originating from highly complex systems is hardly captured by statistics alone. Several approaches to extract the information contained in time-series have been proposed\cite{donner11}. Network analysis based on recurrence and on the analysis of cycles has been employed to study the transition between stable and unstable states in turbulent combustion \cite{godavarthi17,praveen2019,tandon21,gotoda2,gotoda3}. The visibility graph, which maps the steps of a time-series into the nodes of a network whose connection is determined on the basis of mutual visibility\cite{lacasa2008}, has been employed in the analysis of turbulence, especially for fully developed channel, boundary layer flows and jets\cite{iacobello2018,iacobello2019,chowdhuri2021,iacobello2021a,gotoda1}.

In this work, we apply the visibility graph to time-series extracted from a numerically simulated transitional flat plate boundary layer. 
\correct{The visibility graph retains the underlying structure of the process generating the time-series itself and highlights the presence of key patterns retained by it. Moreover, the information contained inside the visibility graph can be condensed into scalar metrics using tools derived from network theory, that we will show to be sensitive to the features of time-series. Thus, we are able to identify the key elements that precede and trigger transition, and to provide a thorough description of the spatial evolution of transition.}

The paper is organized as follows. In Section \ref{sec:methods} the numerical simulation is detailed, the visibility graph method and the relevant network measures are introduced, and the properties of time-series are correlated with those of the visibility graph by means of the parametric analysis of a synthetic time-series. In Section \ref{sec:results} the results are detailed, including a discussion on the application of the present method to badly resolved data. Finally, in Section \ref{sec:concl} concluding remarks are given.

\section{Methods}
\label{sec:methods}

\subsection{Transitional boundary layer dataset description}

\begin{figure}[ht!]
\centering
\includegraphics[width = .45\textwidth]{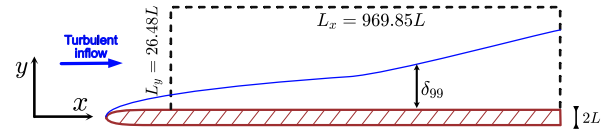}
\caption{Sketch of the fluid domain. The region bounded by the dashed line indicates the stored portion of the domain.\label{fig:new}}
\end{figure}

We apply the visibility graph analysis to velocity time-series extracted from a numerically simulated transitional boundary layer. The velocity fields have been made available through the John Hopkins Turbulence Database. The data is obtained via a direct numerical simulation of the flow over a flat plate of thickness $2L$ with an elliptical leading edge and a zero mean pressure gradient across the streamwise $x$ direction. The origin ($x = 0$) of the domain is located at the leading edge of the plate; this location is excluded from the stored dataset, which instead starts at approximately $x = 30L$. A sketch of the simulation domain is shown in figure \ref{fig:new}. The Reynolds number $Re_L$ based on the half-thickness of the plate, the free-stream velocity $U_{\infty}$ and the fluid viscosity $\nu$ is $Re_L = U_{\infty}L/\nu = 800$.
The flow at the inlet is fully turbulent and is generated from a distinct simulation of homogeneous turbulence. The turbulence intensity at the inlet is about 3\%, which is enough to trigger bypass transition. At the lower boundary, which is a solid wall, the no-slip condition is imposed, while along the spanwise direction periodicity is used. At the top of the domain the boundary condition is actively controlled to satisfy continuity and keep the zero pressure gradient.
The size of the domain, with respect to the plate half-thickness is $L_x \times L_y \times L_z = \left(969.8465 \times 26.4844 \times 240\right) L$, with $y$ and $z$ being the wall-normal and spanwise directions, respectively.
The number of grid points in physical space at which the solution is stored is $N_x \times N_y \times N_z = 3320\times 224\times 2048$. Each time series is composed of $N_t = 4701$ time-steps for a total time stored $T = 1175 L/U_{\infty}$; the resulting time step is $\Delta t = 0.25 L/U_{\infty}$.
A snapshot of the streamwise velocity field at $y=0.43$ is shown in figure \ref{fig:1}(a).
Further details on the numerical procedure employed can be found elsewhere\cite{zaki2013ftc, jhtdb1, jhtdb2}.

\subsection{Visibility graph and network measures}

\begin{figure*}[bt!]
\centering
\includegraphics[width = \textwidth]{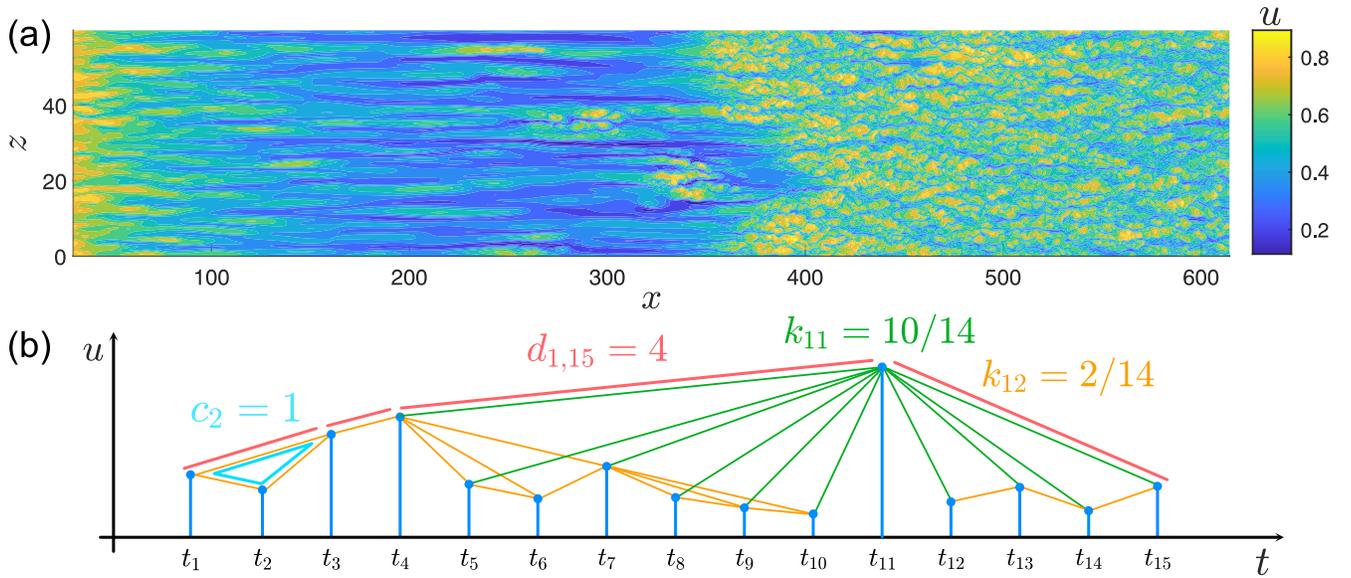}
\caption{(a) top view of the boundary layer at $y = 0.48$, $t = 0$, streamwise velocity; (b) construction of the visibility graph from a time-series and network measures. \correct{Blue points mark the sampled velocity values $u(t_i)$ and orange segments highlight the visibility links between pairs of connected nodes. Links connected to the node $t_{11}$ are colored in green, to show an example of the degree centrality $k_{11}$, while the path between node $t_{1}$ and node $t_{15}$ is highlighted in red, to highlight an example of a shortest path, $d_{1,15}$; finally, the links of the subgraph induced by nodes $t_1$, $t_2$ and $t_3$ is shown in light blue color, to show an example of the local clustering coefficient $c_2$}.\label{fig:1}}
\end{figure*}

The natural visibility graph maps a discrete time-series (or an univariate function, in general) into a graph \cite{lacasa2008}. A graph $\mathcal{G}\left(\mathcal{N},\mathcal{E}\right)$ is an object comprising of a set of nodes $\mathcal{N} = \lbrace n_1,\ldots,n_N \rbrace$ and a set of links $\mathcal{E}$, each one connecting a pair of nodes of the graph. A common representation of a graph is the adjacency matrix $\mathbf{A}\in \mathbb{R}^{N\times N}$, whose entries $A_{ij}$ are equal to one if there is a connection between nodes $n_i$ and $n_j$ and is zero otherwise \cite{boccaletti2006}. 
The visibility method transforms a time-series $u(t_i)$ into a graph by assigning each time step $t_i$ to a node and establishing a link if there is a direct line of sight between the two nodes. More specifically, two nodes $i$ and $j$ corresponding to time steps $t_i$ and $t_j$ are connected if
\begin{equation}
\label{eq:viscrit}
u(t_k) \leqslant \left(u(t_j) - u(t_i)\right)\frac{t_k - t_i}{t_j-t_i},\,\,\, \forall k = i\ldots j
\end{equation}
where $i<j$ without loss of generality. The procedure by which links are formed between nodes is shown schematically in figure \ref{fig:1}(b). \correct{Geometrically, this corresponds to creating a link between two nodes if and only if an uninterrupted straight line can be traced between the corresponding data points, without intersecting any intermediate point of the time-series. This criterion is found to preserve the properties of the time-series and to translate its features into recognizable topological structures contained in the visibility graph \cite{lacasa2008}.}

The resulting graph is fully connected (there is a path between all pair of nodes), undirected (thus having a symmetric adjacency matrix) and has a number $N_t$ of nodes, equal to the number of steps in the time-series.
The visibility graph is invariant to scale transformations of the time-series, since its links are defined following only a convexity criterion. Moreover, unlike methods such as recurrence networks, the visibility approach does not rely on parameters set by the user (such as the phase-space threshold distance in recurrence networks).

Network science adopts several metrics that, when applied to complex graphs (\textit{i.e.} graphs with a large number of nodes and a nontrivial interconnection pattern), give a quick glance on the network properties. We will now enumerate the relevant network measures employed in this work and, subsequently, explain their relevance in the context of visibility analysis.

The \textit{degree centrality} $k_i$ of a node is the number of links incident to that node. Using the adjacency matrix, the degree normalized by its maximum attainable value (that is the number of nodes $N$) is
\begin{equation}
\label{eq:deg}
k_i = \sum_{j = 1}^{N} \frac{A_{ij}}{N-1}
\end{equation}
The degree centrality is one of the most straightforward measures of centrality of a node, \textit{i.e.} of its importance in the overall network structure. In figure \ref{fig:1}(b), the degree centrality of two nodes is computed.

The clustering coefficient $c_i$ is, instead, a measure of the local \textit{density} of links around a node. More specifically, it expresses the probability that two neighbors of node $i$ are themselves connected, thus that a connected triple of nodes is also a triangle \cite{watts1998}. It can be calculated as
\begin{equation}
\label{eq:cc}
c_i = \frac{\sum_{j,m} A_{ij}A_{jm}A_{mi}}{k_i\left(k_i-1\right)}.
\end{equation}
In figure \ref{fig:1}(b) the computation of the clustering coefficient of the second node, $c_2$, is exemplified; as its immediate neighbors form a complete graph, its clustering coefficient is equal to 1.

In order to measure the likeliness of connected nodes, it may be useful to consider the similarity of certain properties of each node of a link, \textit{i.e.} the \textit{assortativity} of the network. In its most basic form, the assortativity $r$ can be measured as the Pearson correlation coefficient of the degree of the two nodes at the ends of a link, computed for all connected node pairs \cite{newman2002,newman2003}. The value of $r$ ranges from -1, indicating a fully disassortative network, to 1 for fully assortative networks in which connections take place only between similar (with respect to the degree centrality) nodes. As an example, nodes 11 and 12 in figure \ref{fig:1}(b) \correct{(which are marked by their respective time steps, $t_{11}$ and $t_{12}$)} have largely different degree centralities, thus they contribute to make the network disassortative.

Finally, the average (or characteristic) path length $L$ is the average length of the shortest paths connecting all pair of nodes, that is
\begin{equation}
\label{eq:avgpl}
L = \frac{1}{N(N-1)}\sum_{i,j,i\neq j} d_{ij},
\end{equation}
where $d_{ij}$ is the topological length of the shortest path between nodes $i$ and $j$. The average path length is a measure of the typical distance between nodes in a graph and can be used as a way to determine the effectiveness of a network in transferring information between nodes. In figure \ref{fig:1}(b) the shortest path between nodes 1 and 15 is made of four links, so that $d_{1,15} = 4$; in this case, the presence of a clear peak in the time-series makes topological connection between temporally distant nodes shorter. We note that if disconnected components are present in the network, the average path length diverges; as the visibility graph is fully connected, in our case $L$ is always bounded.

\begin{figure*}
\centering
\includegraphics[width = .95\textwidth]{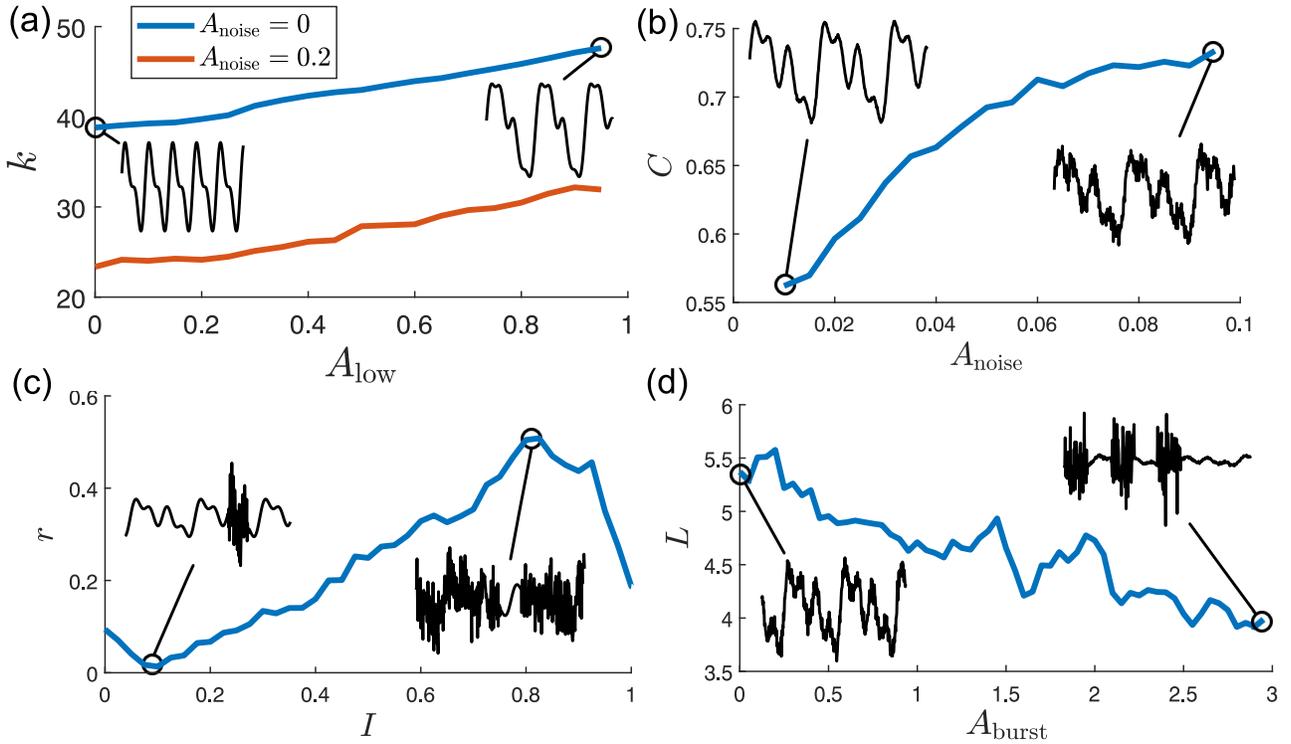}
\caption{Visibility parameters of the synthetic time-series: a) average degree $K$ with respect to the amplitude of the lowest frequency component of the time-series $A_{\mathrm{low}}$; b) clustering coefficient $C$ with respect to the amplitude of noise $A_{\mathrm{noise}}$; c) assortativity $r$ with respect to the intermittency $I$; d) average path length $L$ with respect to the amplitude $A_{\mathrm{burst}}$ of localized bursts ($I = 0.33$). Relevant time-series are plotted as insets. \label{fig:2}}
\end{figure*}

\subsection{Visibility analysis of time-series}
The application of the visibility graph formalism to time-series is known to preserve their structure, in the sense that a relation between time-series properties and those of the visibility graph can be found. As an example, periodic time-series result in regular visibility graphs, random series in random graphs, and fractal series result in graphs where scale-free features are present \cite{lacasa2008}. More recently, and with application to fluid dynamics, the relation between the time-series structure and network measures (as those presented before) has been explored in the context of the visibility graph \cite{iacobello2018, iacobello2019, iacobello2021a, chowdhuri2021, murugesan2015, singh2017}. The main rationale behind these approaches stems from the need to provide a treatable insight on turbulent or transitional time-series, which are highly complex and need a very fine temporal resolution to be adequately represented (especially at high Reynolds numbers).
By construction, the visibility graph is invariant under horizontal and vertical rescaling of the time-series; as such, the overall amplitude of the time-series has no effect on the derived network measures. On the contrary, the visibility analysis is highly sensitive to the interplay of different scales inside the time-series. As will be detailed in the following, the network measures applied to the visibility graph have the ability to convey the entity of this interplay. Finally, the visibility graph can be computed from a time-series with fast algorithms (in $\mathcal{O}(N\log N)$ time, $N$ being the number of nodes\cite{lan2015}), which makes it suitable for use in large datasets.

To establish the properties of the visibility analysis, we now aim to show the behavior of network measures in response to the features of the time-series. To do so, we synthetically generated a time-series by superimposing three out of phase sinusoidal components of different frequency $f_i$ and amplitude $\phi_i$, a small-scale Gaussian noise and a larger amplitude Gaussian noise that has a discontinuous support (in order to mimic the intermittent behavior typical of transitional time-series). The synthetic time-series has the expression
\begin{equation}
u(t) = \sum_{i = \mathrm{low}, \mathrm{mid}, \mathrm{high}} A_i \sin\left(f_i t + \phi_i\right) + A_{\mathrm{noise}}w_1(t) + A_{\mathrm{burst}}w_2(t),
\end{equation}
where $w_1$ and $w_2$ are white Gaussian noises and $w_2$ is nonzero only on a fraction $I$ of the duration of the time-series, \correct{which corresponds to the intermittency of the signal}. The support for the intermittent regions of the synthetic time-series is chosen at random.

In a visibility graph, the nodes situated in large, convex portion of the time-series have a direct line of sight with a larger number of other nodes, thus having large degree. Overall, the mean degree 
\begin{equation}
K = \frac{1}{N} \sum_{i = 1}^{N} k_{i} = \frac{1}{N(N-1)} \sum_{i,j} A_{ij}
\end{equation}
of the visibility graph is tightly connected to the amplitude $A_{\mathrm{low}}$ of the lowest frequency components in the time-series, \textit{i.e.}, those with the largest scale. All the panels in figure \ref{fig:2} show some of the synthetic time-series used to compute the visibility graph measures.
Figure \ref{fig:2}(a) shows how the degree grows as the low-frequency component becomes more important. Indeed, the preminence of peaks spaced far apart increases the overall number of links and thus the mean degree; also, we found that the mean degree is inversely proportional to the frequency $f_{\mathrm{low}}$.

The clustering coefficient also depends strongly on the local convexity of the time-series around a given node. Differently from the degree centrality, the amount of connected triples and triangles is mostly determined by the time-steps immediately adjacent to the one considered, as it is far more frequent to find connections in triples of temporally close nodes. As such, the global average 
\begin{equation}
C = \frac{1}{N} \sum_{i = 1}^{N} c_{i} \in \left[0,\,1\right]
\end{equation}
of the clustering coefficient quantifies the importance of the small-scale components of the time-series \citep{iacobello2018}. In particular, we observe that the clustering coefficient increases as the amplitude of the small-scale Gaussian noise $A_{\mathrm{noise}}$ in the synthetic time-series increases (see figure \ref{fig:2}(b)).

The visibility graph is also suited towards the analysis of the vertical separation, \textit{i.e.}, the presence of components whose amplitude is markedly different in adjacent regions of the time-series. \textcite{iacobello2019} gave an extended discussion regarding the ability of the assortativity coefficient $r$ to discern between time-series whose amplitude is homogeneous over time and those that are not. In particular, when the amplitude of the time-series is homogeneous and there are no outliers, it is more probable that similar nodes are connected, leading to highly assortative networks. This behavior is of outstanding importance in the study of transitional time-series, which are strongly characterized by their intermittent behavior in the region where turbulence is not fully developed.
Varying the intermittence in the synthetic time-series (that is the percentage of time, chosen at random, in which large-amplitude Gaussian noise is present) allowed us to study the behavior of the assortativity, as is shown in figure \ref{fig:2}(c). The "turbulent" component is highly fluctuating, but is still somewhat homogeneous with respect to the underlying laminar component; accordingly, the assortativity is high. As the intermittency grows the network becomes increasingly assortative until the intermittency reaches unity (the series becomes fully turbulent), where the assortativity drops significantly. As already stated, the visibility graph is highly sensitive to the interplay of the different scales rather than to their actual amplitude. When the intermittency reaches unity, the laminar component disappears from the time-series and the heterogeneity of the turbulent component leads to a decrease of the assortativity. 

Finally, the average path length is a measure of the topological distance between nodes; it follows easily from this consideration that the presence of localized peaks determines a decrease of the value of $L$, as long-distance nodes are more easily connected. As shown in figure \ref{fig:2}(d), the average path length decreases when the amplitude of the temporally localized Gaussian noise in the synthetic time-series increases. 

\section{Results}
\label{sec:results}
\subsection{Network measures}

\begin{figure*}[th!]
\centering
\includegraphics[width = .95\textwidth]{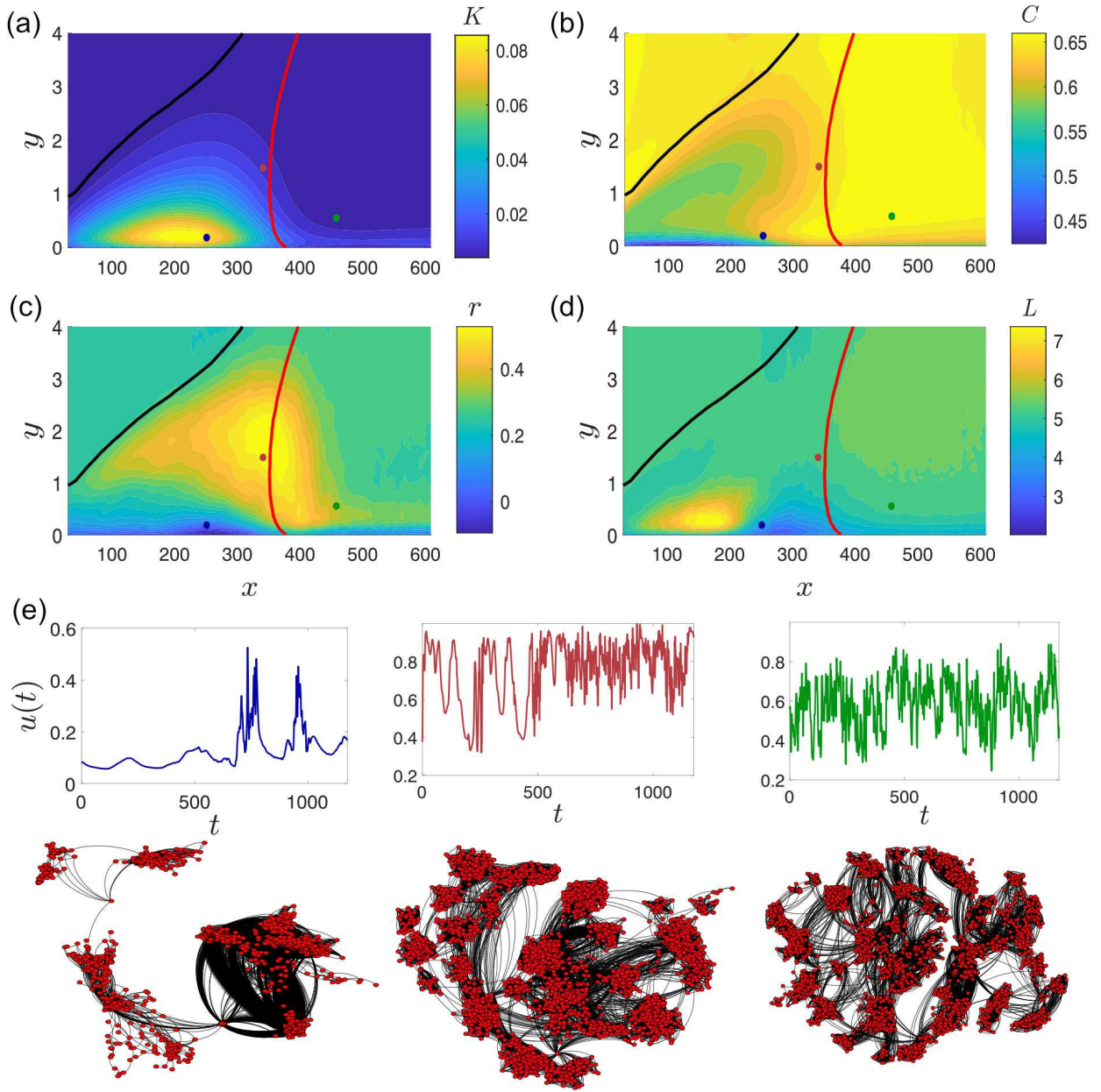}
\caption{Visibility graph measures in the transitional boundary layer: a) Average degree, b) clustering coefficient, c) assortativity and d) average path length. The boundary layer height $\delta_{99}$ is shown in black, while the average position of the TNTI $\bar{x}_{\mathrm{TNTI}}(y)$ is shown in red. Panel e) shows three time-series sampled at the corresponding, color-coded points of the domain; the spanwise coordinate is $z = 13.95$ for all time-series. Below each time-series, the corresponding visibility graph is drawn by a force-directed algorithm. \label{fig:3}}
\end{figure*}

We computed the visibility graph from time-series of the streamwise velocity $u$ from a set of points in the flow domain. In particular, for each $(x,\, y)$ coordinate, we considered time-series from $N_z = 120$ equally spaced points across the homogeneous $z$ direction of the domain. For each visibility graph, we computed the relevant network measures; in the following, we will provide results averaged along $z$.

Figures \ref{fig:2}(a)-(d) show the four visibility graph measures ($K$, $C$, $r$ and $L$) over a grid of $(x, y)$ points that encompass the regions of the boundary layer where bypass transition takes place. Taking into account the significance of network measures in the context of the visibility graph, some features of the boundary layer can be readily identified.
As stated before, the turbulent fluctuations in the boundary layer induces low-frequency, high-amplitude fluctuations of the streamwise velocity in the inner portion of the boundary layer. The degree centrality (figure \ref{fig:3}(a)), which is sensitive to these low-frequency components, has a marked peak near $(x,\,y) = (200,\,0.2)L$, where the low-frequency amplification induced by shear-sheltering appears to be at a maximum. As transition spatially progresses, the low frequency streaks encounter secondary instabilities which effectively enable the transfer of energy to smaller scales, initiating the breakdown to turbulence. Using the clustering coefficient (figure \ref{fig:3}(b)), we are able to locate the start of this process at around $x = 300L$ in the innermost portion of the boundary layer, while at higher $y$ values the high-frequency components of the time-series acquire stronger importance at higher $x$. Transition appears to be initiated in the innermost portion of the boundary layer. We also note that, in the region where streaks are generated, the clustering coefficient is at a minimum, as the high-frequency components of the freestream turbulence are filtered out and do not exert forcing on the boundary layer flow. Additionally, the clustering coefficient in the turbulent region of the boundary layer is slightly higher than in the free-stream, indicating some difference in the structure of turbulence (regardless of the amplitude, which is neglected by the visibility graph).

The assortativity (figure \ref{fig:3}(c)) and the average path length (figure \ref{fig:3}(d)) allow us to investigate the intermittent behavior of the time-series just before full transition occurs. As already stated, these time-series are characterized by the passage, in fixed points of the domain, of turbulent spots. As $x$ grows, the size and frequency of spots grows and, accordingly, the intermittency of the time-series grows. The remarkable ability of the assortativity to distinguish between intermittent time-series (even with high intermittency values) and fully turbulent ones allows us to identify a region, located prior to the fully developed turbulent boundary layer, where intermittency is at a maximum (but still not unity). This region with high assortativity presents itself as an almost vertical front located at around $x = 380L$, which is followed by a decrease in assortativity. We hypothesize that the region of maximum assortativity is the region which hosts the turbulent-non turbulent interface. 
The average path length has a maximum in the streaky region, where the low-frequency fluctuations of the streamwise velocity present no clear, localized peak. Instead, there is a minimum at around $x = 300L$ and very low $y$ values. The minimum indicates that time-series in this region present localized peaks and a somewhat reduced intermittency (the assortativity is also very low in the same region, indicating a strongly heterogeneous time-series). It is interesting to note that at slightly higher $x$ coordinates the strong increase of the clustering coefficient takes place; indeed, it is the expansion and coalescence of localized spots that generates complete transition to turbulence.

Figure \ref{fig:3}(e) shows three time-series extracted from the corresponding, color-coded, points indicated in figures \ref{fig:3}(a)-(d) and at $z = 13.95$ and, below, the corresponding visibility graph plotted with a force-directed algorithm\cite{openord}. The leftmost time-series is taken from the region of the domain in which turbulent spots are present. 
Here the presence of peaks, which are also evident in the graph plot as single nodes connected with large, separate clusters, contributes to the low value of $L$, while the presence of laminar regions determines the high value of $K$. 
\correct{To better identify these large clusters we applied the Louvain partitioning algorithm\cite{louvain} and computed the average size of communities. In the case of the leftmost time-series, the average cluster comprises 522 nodes.}
In the central time-series and in its corresponding graph plot, taken near the turbulent-non turbulent interface where the \correct{assortativity} $r$ is at a maximum, the hub-spoke organization typical of the spotty time-series is less evident, but clusters still appear visibly larger than in the fully turbulent case due to the contribution of persistent low frequency components \correct{(the average cluster size is 204)}; consequently, the degree is also slightly larger with respect to that of the fully turbulent time-series. Finally, in the rightmost time-series, taken from a region where turbulence is fully developed, the size of clusters is much smaller \correct{(on average, they comprise 130 nodes)}. Conversely, here the clustering coefficient is at a maximum as the high-frequency components are predominant.

To further complement our results, we added to the plots in figure \ref{fig:3} the boundary layer height $\delta_{99}$ (black curves), which is the $y$ coordinate at which the streamwise average velocity is 99\% the freestream velocity, and an approximate location of the TNTI (red curves). To find the location of the turbulent-non turbulent interface across the wall-normal coordinate $y$ we used the procedure proposed by \textcite{nolan2013}.
In particular, for each time step of the boundary layer simulation data and at each wall-normal height, we applied Otsu's method\cite{otsu1979} to the sum of the wall-normal and spanwise velocity fluctuations $\lvert v'\rvert + \lvert w'\rvert$ in order to obtain a threshold value (as a function of $y$ and $t$). Otsu's method is an image segmentation technique that identifies the optimal threshold of a scalar value (the grayscale in images or $\lvert v'\rvert + \lvert w'\rvert$ in the boundary layer). In particular, the method achieves the objective of minimizing the \textit{intra-class} variance within the turbulent and the laminar regions, and does so by maximizing the \textit{inter-class} variance between the laminar and turbulent regions. Because of the dependence on $y$ of the wall-normal velocity fluctuations, a global ($y$-independent) threshold cannot be provided and Otsu's procedure has to be applied to isolated vertical slices of the boundary layer flow, as prescribed by \textcite{nolan2013}.
After we obtained the spatial profile of the TNTI as a surface $x_{\mathrm{TNTI}}(y,z,t)$ at different times we averaged the results along time and along $z$ to obtain an expected average profile of the TNTI, $\bar{x}_{\mathrm{TNTI}}(y)$. 

First, we note that the variation of the network measures takes place inside the boundary layer, at $y$ coordinates lower than $\delta_{99}$, while in the free-stream the measures remain mostly stationary. Moreover, the expected position of the turbulent-non turbulent interface using Otsu's thresholding is clearly superimposable to the regions where network measures signal the transition to fully developed turbulence. In particular, the region of maximum assortativity and the TNTI intersect in a wide range of $y$ coordinates, indicating that the assortativity measures, as was hypothesized before, incipient transition. It is also worth noting that the spike of the clustering coefficient in the innermost region of the boundary layer is located slightly before the expected location of the interface, possibly indicating that to some extent the transfer of energy to smaller scales precedes full transition.

\subsection{Sensitivity analysis}

\begin{figure*}
\centering
\includegraphics[width = .96\textwidth]{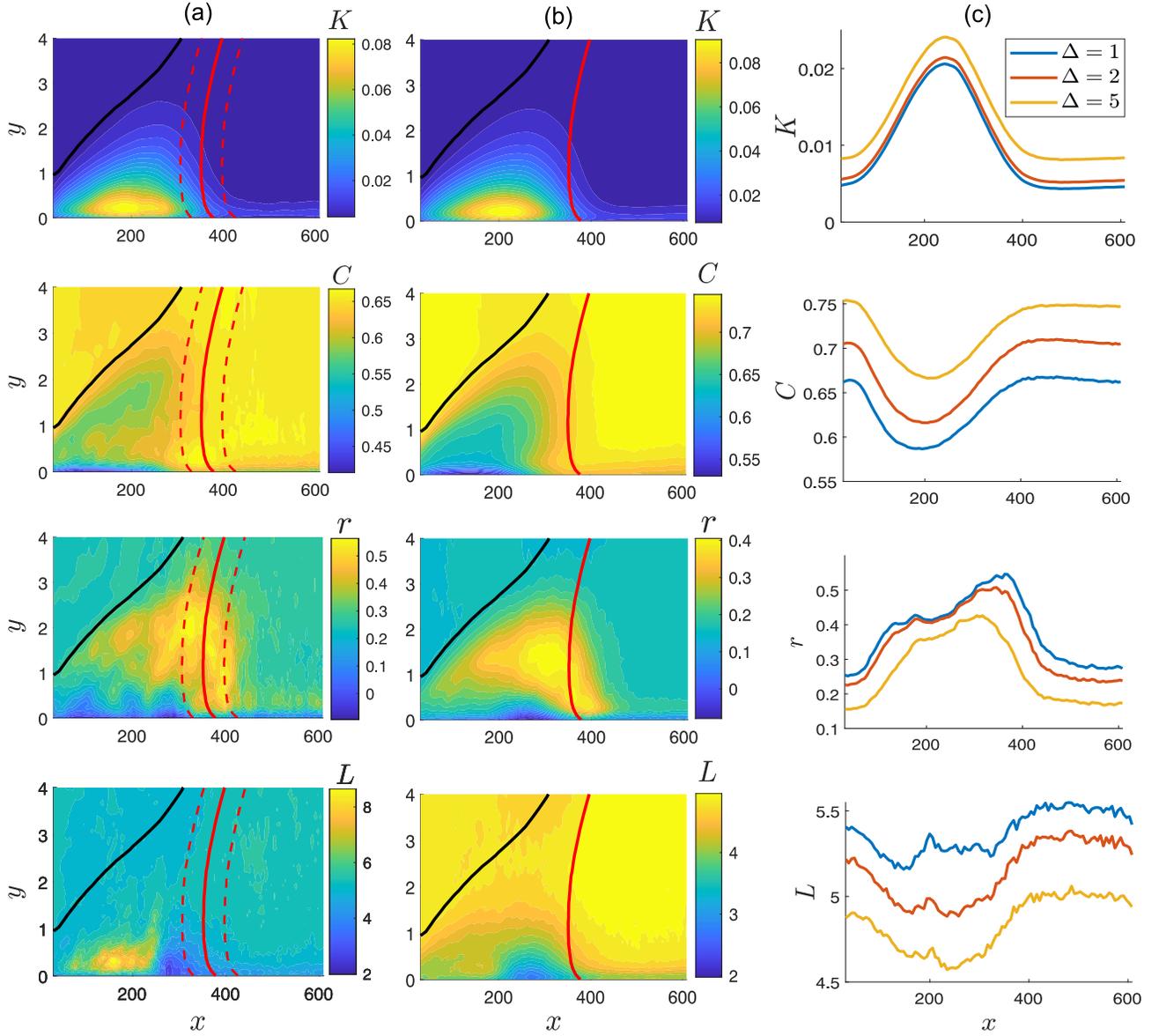}
\caption{a) Degree, clustering coefficient, assortativity and average path length of the visibility graph at a fixed spanwise coordinate ($z = 13.95$); black lines indicate the values of $\delta_{99}$, while the red lines the average position of the TNTI and the dashed red lines its standard deviation along $z$. b) Measures for networks obtained from subsampled time-series with $\Delta = 5$ (results are now averaged along $N_z = 120$ points in the spanwise direction). c) Network measures from the subsampled time-series at $y = 0.98$ with different values of the subsampling parameter $\Delta$ (again, results are averaged along $N_z = 120$ points)\label{fig:4}}
\end{figure*}

We now aim to assess the response of the visibility analysis to the decrease of spatial and temporal resolution. In particular, we will provide results that are averaged along a reduced number of points along $z$ and results obtained using subsampled time-series. The analysis of subsampled time-series is relevant because it may provide useful insight in the analysis of data with lower resolution, such as experiments having sparser data acquisition setup or LES simulations over complex 3D geometries.

Panels from figure \ref{fig:4}(a) show the network measures averaged using only $10$ points along $z$ (instead of $N_z = 120$ used precedently); we also plotted the location of the TNTI computed using Otsu's method and its standard deviation along $z$ (dashed lines). The overall trends of the measures across the domain are preserved. In particular, a peak of the assortativity is still present near the transition region, making this measure suitable for a local determination of the TNTI. Network properties do not change significantly along the homogeneous direction of the domain. 

Figures \ref{fig:4}(b)-(c) show the network measures obtained from subsampled time-series. Starting from the full time-series, comprising of $N_t$ time-steps, we constructed the subsampled time-series by taking one out of $\Delta$ points, so that $u^{\Delta}_i = u_{\Delta i}$. We thus obtained subsampled time-series which are effectively deprived of the highest frequency components. The effect of the subsampling from the perspective of the visibility graph is a result of both the elimination of a fraction of the original data and of the reduction in the number of nodes.

In figure \ref{fig:4}(b), the four network measures ($K$, $C$, $r$ and $L$) obtained from a subsampled time-series with $\Delta = 5$ are shown. The degree centrality has been normalized with the correct number of nodes, which is $N_t/\Delta$. Even if 80\% of the original information contained in the time-series is lost, the visibility measures of the subsampled time-series are tightly related to those of the full ones. All the previously identified spatial patterns are preserved through the subsampling. The degree centrality is left mostly unchanged, as the subsampling procedure affects only slightly the larger timescale in the time-series (which are the ones that influence the degree centrality); moreover, the decrease of the number of nodes can be fully accounted for by normalization. The clustering coefficient, while retaining the previously found spatial trends (most notably, the presence of a minimum around $x = 200$), increases everywhere in the domain. While the removal of the high-frequency components triggers a decrease of the clustering coefficient at a fixed number of time-steps (see figure \ref{fig:2}(b)), this seems to not be the case when the number of time-steps is decreased. Moreover, it appears that the increase of the clustering coefficient has the same magnitude in regions of the domain where the time-series are dominated by either low- or high-frequency components, indicating that the reduction of the number of nodes has the most prominent effect on the visibility graph structure.
The assortativity appears to be slightly reduced everywhere, although not by much. Moreover, the spatial location of its maximum, which we correlated to the location of the TNTI, appears to be located at slightly lower $x$ coordinates. We hypothesize that this is due to the loss of information occurring because of the subsampling, which makes it impossible to distinguish a time-series with a high value of the intermittency $I$ from a fully turbulent one.
Finally, the average path length $L$ diminishes everywhere as the lesser number of nodes is correlated to a reduction of the shortest path lengths. Around $x = 200$ the average path length of the full time-series presents a peak, which is progressively smoothed as the subsampling parameter $\Delta$ increases. In that region of the domain the dynamics of the flow is dominated by low-frequency streaks and turbulent spots are mostly yet to appear, which, in the fully sampled case, leads to high values of $L$. It appears that fine-structure changes like those induced by the subsampling considerably affect the structure of the visibility graph obtained from time-series of streaky flow.

The panels of figure \ref{fig:4}(c) show the network measures at a fixed wall-normal coordinate ($y = 0.98$) and two different subsampling parameters, $\Delta = 2$ and 5. With $\Delta = 2$, even if half the information of the original time-series is lost, the network properties behave in a similar manner to the fully sampled ones. In particular, we note that the assortativity $r$ is mostly unchanged and the peak is located at the same $x$-location as in the fully sampled case. Moreover, a peak of $L$ in the streaky region is still present, indicating that the effects of the subsampling with $\Delta = 2$ are mostly due to the halving of the number of nodes. 

\correct{We performed the same analysis at higher values of $\Delta$ (up to $\Delta = 20$) obtaining a progressive loss of quality in the network measures. This is indeed expected, as the loss of information due to the subsampling inevitably reflects on the quality of the visibility graph analysis. Nonetheless, if the sampling guarantees that all the relevant scales of the flow are retained, the qualitative behavior of the results obtained through the visibility graph is independent from the exact value of the sampling rate.}

\section{Conclusions}
\label{sec:concl}

We performed a visibility analysis on time-series obtained from a numerically simulated transitional boundary layer. After characterizing the behavior of network measures using a parametric time-series, we computed these measures for the visibility graphs obtained from streamwise velocity time-series extracted from the flow domain. The four metrics, namely the degree centrality $K$, the clustering coefficient $C$, the assortativity $r$ and the average path length $L$ are together able to accurately provide a view on the spatial evolution of the bypass transition from laminar to turbulent flow, as they encode the interplay of scales and the transformations occurring due to the onset of turbulence.

Most notably, we found that the assortativity $r$, \textit{i.e.} the Pearson correlation coefficient between the degree of linked pair of nodes, has a peak in the region of the domain immediately preceding the rise of developed turbulence and is thus able to act as a reliable onset marker for transition. This finding is noteworthy since the visibility analysis does not require any \textit{a priori} parameter and, most importantly, because an extensive knowledge of the velocity field is not needed. Indeed, our results are obtained from the streamwise component $u$ of the velocity and from single point measurements only. To identify a time-averaged location of the TNTI using the present approach, one would only need to compute the assortativity coefficient of visibility graph from different $x$ coordinates at a fixed wall-normal height $y$ and spanwise coordinate $z$ and find the location of the peak of $r$.
Our approach provides results in agreement with more established methods, such as those based on thresholding physical quantities through Otsu's algorithm, but does not need the same extensive knowledge of the flow field of these methods. Although the thresholding of the indicator function $\lvert v' \rvert + \lvert w' \rvert$ provides a time-instantaneous location of the TNTI, it requires the knowledge of the velocity field on at least a slab of the domain at constant wall-normal height with reasonable resolution.
Moreover, we found that similar results are obtained using subsampled time-series, confirming the robustness of the method with incomplete data.

\correct{The visibility graph methodology appears suited towards the analysis of flows with abrupt changes in their dynamics. Accordingly, the current approach can be applied in future works to investigate the influence of key factors influencing transition to turbulence (such as geometry, wall roughness, pressure gradient). Moreover, the visibility approach could be extended to respect temporal causality by including a directional information into the link definition, thus creating a directed graph. Past research has shown how this allows the visibility graph to detect time-series generated by irreversible process and could be useful to identify and quantify the arising of irreversibility due to transition to turbulence \cite{lacasa2012time}.}
Overall, network-based methods have an interesting outlook with regard to application in fluid dynamics, as they are able to capture the interplay of scales typical of turbulent, highly complex flows and provide meaningful and consistent results when applied to highly different cases, such as the laminar, intermittent and turbulent portions of a transitional boundary layer flow.

\begin{acknowledgments}
Computational resources were provided by HPC@POLITO, a project of Academic Computing within the Department of Control and Computer Engineering at the Politecnico di Torino (\url{https://www.hpc.polito.it})
\end{acknowledgments}

\nocite{*}
\bibliography{aipsamp}

\providecommand{\noopsort}[1]{}\providecommand{\singleletter}[1]{#1}%
\begin{thebibliography}{46}%
\makeatletter
\providecommand \@ifxundefined [1]{%
 \@ifx{#1\undefined}
}%
\providecommand \@ifnum [1]{%
 \ifnum #1\expandafter \@firstoftwo
 \else \expandafter \@secondoftwo
 \fi
}%
\providecommand \@ifx [1]{%
 \ifx #1\expandafter \@firstoftwo
 \else \expandafter \@secondoftwo
 \fi
}%
\providecommand \natexlab [1]{#1}%
\providecommand \enquote  [1]{``#1''}%
\providecommand \bibnamefont  [1]{#1}%
\providecommand \bibfnamefont [1]{#1}%
\providecommand \citenamefont [1]{#1}%
\providecommand \href@noop [0]{\@secondoftwo}%
\providecommand \href [0]{\begingroup \@sanitize@url \@href}%
\providecommand \@href[1]{\@@startlink{#1}\@@href}%
\providecommand \@@href[1]{\endgroup#1\@@endlink}%
\providecommand \@sanitize@url [0]{\catcode `\\12\catcode `\$12\catcode
  `\&12\catcode `\#12\catcode `\^12\catcode `\_12\catcode `\%12\relax}%
\providecommand \@@startlink[1]{}%
\providecommand \@@endlink[0]{}%
\providecommand \url  [0]{\begingroup\@sanitize@url \@url }%
\providecommand \@url [1]{\endgroup\@href {#1}{\urlprefix }}%
\providecommand \urlprefix  [0]{URL }%
\providecommand \Eprint [0]{\href }%
\providecommand \doibase [0]{https://doi.org/}%
\providecommand \selectlanguage [0]{\@gobble}%
\providecommand \bibinfo  [0]{\@secondoftwo}%
\providecommand \bibfield  [0]{\@secondoftwo}%
\providecommand \translation [1]{[#1]}%
\providecommand \BibitemOpen [0]{}%
\providecommand \bibitemStop [0]{}%
\providecommand \bibitemNoStop [0]{.\EOS\space}%
\providecommand \EOS [0]{\spacefactor3000\relax}%
\providecommand \BibitemShut  [1]{\csname bibitem#1\endcsname}%
\let\auto@bib@innerbib\@empty
\bibitem [{\citenamefont {Zaki}(2013)}]{zaki2013ftc}%
  \BibitemOpen
  \bibfield  {author} {\bibinfo {author} {\bibfnamefont {T.~A.}\ \bibnamefont
  {Zaki}},\ }\bibfield  {title} {\enquote {\bibinfo {title} {From streaks to
  spots and on to turbulence: Exploring the dynamics of boundary layer
  transition},}\ }\href {https://doi.org/10.1007/s10494-013-9502-8} {\bibfield
  {journal} {\bibinfo  {journal} {Flow, Turbulence and Combustion}\ }\textbf
  {\bibinfo {volume} {3}},\ \bibinfo {pages} {451--473} (\bibinfo {year}
  {2013})}\BibitemShut {NoStop}%
\bibitem [{\citenamefont {Vermeersch}\ and\ \citenamefont
  {Arnal}(2010)}]{vermeersch2010}%
  \BibitemOpen
  \bibfield  {author} {\bibinfo {author} {\bibfnamefont {O.}~\bibnamefont
  {Vermeersch}}\ and\ \bibinfo {author} {\bibfnamefont {D.}~\bibnamefont
  {Arnal}},\ }\bibfield  {title} {\enquote {\bibinfo {title} {Klebanoff-mode
  modeling and bypass-transition prediction},}\ }\href
  {https://doi.org/10.2514/1.J050002} {\bibfield  {journal} {\bibinfo
  {journal} {AIAA Journal}\ }\textbf {\bibinfo {volume} {48}},\ \bibinfo
  {pages} {2491--2500} (\bibinfo {year} {2010})},\ \Eprint
  {https://arxiv.org/abs/https://doi.org/10.2514/1.J050002}
  {https://doi.org/10.2514/1.J050002} \BibitemShut {NoStop}%
\bibitem [{\citenamefont {Bhushan}\ \emph {et~al.}(2018)\citenamefont
  {Bhushan}, \citenamefont {Keith~Walters}, \citenamefont {Muthu},\ and\
  \citenamefont {Pasiliao}}]{bhushan2018}%
  \BibitemOpen
  \bibfield  {author} {\bibinfo {author} {\bibfnamefont {S.}~\bibnamefont
  {Bhushan}}, \bibinfo {author} {\bibfnamefont {D.}~\bibnamefont
  {Keith~Walters}}, \bibinfo {author} {\bibfnamefont {S.}~\bibnamefont
  {Muthu}},\ and\ \bibinfo {author} {\bibfnamefont {C.~L.}\ \bibnamefont
  {Pasiliao}},\ }\bibfield  {title} {\enquote {\bibinfo {title}
  {{Identification of Bypass Transition Onset Markers Using Direct Numerical
  Simulation}},}\ }\href {https://doi.org/10.1115/1.4040299} {\bibfield
  {journal} {\bibinfo  {journal} {Journal of Fluids Engineering}\ }\textbf
  {\bibinfo {volume} {140}} (\bibinfo {year} {2018}),\ 10.1115/1.4040299},\
  \bibinfo {note} {111107}\BibitemShut {NoStop}%
\bibitem [{\citenamefont {Sreenivasan}\ and\ \citenamefont
  {Meneveau}(1986)}]{sreenivasan1986}%
  \BibitemOpen
  \bibfield  {author} {\bibinfo {author} {\bibfnamefont {K.~R.}\ \bibnamefont
  {Sreenivasan}}\ and\ \bibinfo {author} {\bibfnamefont {C.}~\bibnamefont
  {Meneveau}},\ }\bibfield  {title} {\enquote {\bibinfo {title} {The fractal
  facets of turbulence},}\ }\href {https://doi.org/10.1017/S0022112086001209}
  {\bibfield  {journal} {\bibinfo  {journal} {Journal of Fluid Mechanics}\
  }\textbf {\bibinfo {volume} {173}},\ \bibinfo {pages} {357--386} (\bibinfo
  {year} {1986})}\BibitemShut {NoStop}%
\bibitem [{\citenamefont {Chauhan}\ \emph {et~al.}(2014)\citenamefont
  {Chauhan}, \citenamefont {Philip}, \citenamefont {de~Silva}, \citenamefont
  {Hutchins},\ and\ \citenamefont {Marusic}}]{chauhan2014}%
  \BibitemOpen
  \bibfield  {author} {\bibinfo {author} {\bibfnamefont {K.}~\bibnamefont
  {Chauhan}}, \bibinfo {author} {\bibfnamefont {J.}~\bibnamefont {Philip}},
  \bibinfo {author} {\bibfnamefont {C.~M.}\ \bibnamefont {de~Silva}}, \bibinfo
  {author} {\bibfnamefont {N.}~\bibnamefont {Hutchins}},\ and\ \bibinfo
  {author} {\bibfnamefont {I.}~\bibnamefont {Marusic}},\ }\bibfield  {title}
  {\enquote {\bibinfo {title} {The turbulent/non-turbulent interface and
  entrainment in a boundary layer},}\ }\href
  {https://doi.org/10.1017/jfm.2013.641} {\bibfield  {journal} {\bibinfo
  {journal} {Journal of Fluid Mechanics}\ }\textbf {\bibinfo {volume} {742}},\
  \bibinfo {pages} {119--151} (\bibinfo {year} {2014})}\BibitemShut {NoStop}%
\bibitem [{\citenamefont {Borrell}\ and\ \citenamefont
  {Jiménez}(2016)}]{borrell2016}%
  \BibitemOpen
  \bibfield  {author} {\bibinfo {author} {\bibfnamefont {G.}~\bibnamefont
  {Borrell}}\ and\ \bibinfo {author} {\bibfnamefont {J.}~\bibnamefont
  {Jiménez}},\ }\bibfield  {title} {\enquote {\bibinfo {title} {Properties of
  the turbulent/non-turbulent interface in boundary layers},}\ }\href
  {https://doi.org/10.1017/jfm.2016.430} {\bibfield  {journal} {\bibinfo
  {journal} {Journal of Fluid Mechanics}\ }\textbf {\bibinfo {volume} {801}},\
  \bibinfo {pages} {554--596} (\bibinfo {year} {2016})}\BibitemShut {NoStop}%
\bibitem [{\citenamefont {Otsu}(1979)}]{otsu1979}%
  \BibitemOpen
  \bibfield  {author} {\bibinfo {author} {\bibfnamefont {N.}~\bibnamefont
  {Otsu}},\ }\bibfield  {title} {\enquote {\bibinfo {title} {A threshold
  selection method from gray-level histograms},}\ }\href
  {https://doi.org/10.1109/TSMC.1979.4310076} {\bibfield  {journal} {\bibinfo
  {journal} {IEEE Transactions on Systems, Man, and Cybernetics}\ }\textbf
  {\bibinfo {volume} {9}},\ \bibinfo {pages} {62--66} (\bibinfo {year}
  {1979})}\BibitemShut {NoStop}%
\bibitem [{\citenamefont {Nolan}\ and\ \citenamefont {Zaki}(2013)}]{nolan2013}%
  \BibitemOpen
  \bibfield  {author} {\bibinfo {author} {\bibfnamefont {K.~P.}\ \bibnamefont
  {Nolan}}\ and\ \bibinfo {author} {\bibfnamefont {T.~A.}\ \bibnamefont
  {Zaki}},\ }\bibfield  {title} {\enquote {\bibinfo {title} {Conditional
  sampling of transitional boundary layers in pressure gradients},}\ }\href
  {https://doi.org/10.1017/jfm.2013.287} {\bibfield  {journal} {\bibinfo
  {journal} {Journal of Fluid Mechanics}\ }\textbf {\bibinfo {volume} {728}},\
  \bibinfo {pages} {306–339} (\bibinfo {year} {2013})}\BibitemShut {NoStop}%
\bibitem [{\citenamefont {Foroozan}\ \emph {et~al.}(2021)\citenamefont
  {Foroozan}, \citenamefont {Guerrero}, \citenamefont {Ianiro},\ and\
  \citenamefont {Discetti}}]{foroozan2021}%
  \BibitemOpen
  \bibfield  {author} {\bibinfo {author} {\bibfnamefont {F.}~\bibnamefont
  {Foroozan}}, \bibinfo {author} {\bibfnamefont {V.}~\bibnamefont {Guerrero}},
  \bibinfo {author} {\bibfnamefont {A.}~\bibnamefont {Ianiro}},\ and\ \bibinfo
  {author} {\bibfnamefont {S.}~\bibnamefont {Discetti}},\ }\bibfield  {title}
  {\enquote {\bibinfo {title} {Unsupervised modelling of a transitional
  boundary layer},}\ }\href {https://doi.org/10.1017/jfm.2021.829} {\bibfield
  {journal} {\bibinfo  {journal} {Journal of Fluid Mechanics}\ }\textbf
  {\bibinfo {volume} {929}},\ \bibinfo {pages} {A3} (\bibinfo {year}
  {2021})}\BibitemShut {NoStop}%
\bibitem [{\citenamefont {Wu}\ \emph {et~al.}(2019)\citenamefont {Wu},
  \citenamefont {Lee}, \citenamefont {Meneveau},\ and\ \citenamefont
  {Zaki}}]{wu2019}%
  \BibitemOpen
  \bibfield  {author} {\bibinfo {author} {\bibfnamefont {Z.}~\bibnamefont
  {Wu}}, \bibinfo {author} {\bibfnamefont {J.}~\bibnamefont {Lee}}, \bibinfo
  {author} {\bibfnamefont {C.}~\bibnamefont {Meneveau}},\ and\ \bibinfo
  {author} {\bibfnamefont {T.}~\bibnamefont {Zaki}},\ }\bibfield  {title}
  {\enquote {\bibinfo {title} {Application of a self-organizing map to identify
  the turbulent-boundary-layer interface in a transitional flow},}\ }\href
  {https://doi.org/10.1103/PhysRevFluids.4.023902} {\bibfield  {journal}
  {\bibinfo  {journal} {Phys. Rev. Fluids}\ }\textbf {\bibinfo {volume} {4}},\
  \bibinfo {pages} {023902} (\bibinfo {year} {2019})}\BibitemShut {NoStop}%
\bibitem [{\citenamefont {Iacobello}, \citenamefont {Ridolfi},\ and\
  \citenamefont {Scarsoglio}(2021{\natexlab{a}})}]{iacobello2021}%
  \BibitemOpen
  \bibfield  {author} {\bibinfo {author} {\bibfnamefont {G.}~\bibnamefont
  {Iacobello}}, \bibinfo {author} {\bibfnamefont {L.}~\bibnamefont {Ridolfi}},\
  and\ \bibinfo {author} {\bibfnamefont {S.}~\bibnamefont {Scarsoglio}},\
  }\bibfield  {title} {\enquote {\bibinfo {title} {A review on turbulent and
  vortical flow analyses via complex networks},}\ }\href
  {https://doi.org/https://doi.org/10.1016/j.physa.2020.125476} {\bibfield
  {journal} {\bibinfo  {journal} {Physica A: Statistical Mechanics and its
  Applications}\ }\textbf {\bibinfo {volume} {563}},\ \bibinfo {pages} {125476}
  (\bibinfo {year} {2021}{\natexlab{a}})}\BibitemShut {NoStop}%
\bibitem [{\citenamefont {Taira}\ and\ \citenamefont {Nair}(2022)}]{taira2022}%
  \BibitemOpen
  \bibfield  {author} {\bibinfo {author} {\bibfnamefont {K.}~\bibnamefont
  {Taira}}\ and\ \bibinfo {author} {\bibfnamefont {A.~G.}\ \bibnamefont
  {Nair}},\ }\bibfield  {title} {\enquote {\bibinfo {title} {Network-based
  analysis of fluid flows: Progress and outlook},}\ }\href
  {https://doi.org/https://doi.org/10.1016/j.paerosci.2022.100823} {\bibfield
  {journal} {\bibinfo  {journal} {Progress in Aerospace Sciences}\ }\textbf
  {\bibinfo {volume} {131}},\ \bibinfo {pages} {100823} (\bibinfo {year}
  {2022})}\BibitemShut {NoStop}%
\bibitem [{\citenamefont {Boccaletti}\ \emph {et~al.}(2006)\citenamefont
  {Boccaletti}, \citenamefont {Latora}, \citenamefont {Moreno}, \citenamefont
  {Chavez},\ and\ \citenamefont {Hwang}}]{boccaletti2006}%
  \BibitemOpen
  \bibfield  {author} {\bibinfo {author} {\bibfnamefont {S.}~\bibnamefont
  {Boccaletti}}, \bibinfo {author} {\bibfnamefont {V.}~\bibnamefont {Latora}},
  \bibinfo {author} {\bibfnamefont {Y.}~\bibnamefont {Moreno}}, \bibinfo
  {author} {\bibfnamefont {M.}~\bibnamefont {Chavez}},\ and\ \bibinfo {author}
  {\bibfnamefont {D.-U.}\ \bibnamefont {Hwang}},\ }\bibfield  {title} {\enquote
  {\bibinfo {title} {Complex networks: Structure and dynamics},}\ }\href
  {https://doi.org/https://doi.org/10.1016/j.physrep.2005.10.009} {\bibfield
  {journal} {\bibinfo  {journal} {Physics Reports}\ }\textbf {\bibinfo {volume}
  {424}},\ \bibinfo {pages} {175--308} (\bibinfo {year} {2006})}\BibitemShut
  {NoStop}%
\bibitem [{\citenamefont {Scarsoglio}, \citenamefont {Iacobello},\ and\
  \citenamefont {Ridolfi}(2016)}]{scarsoglio2016complex}%
  \BibitemOpen
  \bibfield  {author} {\bibinfo {author} {\bibfnamefont {S.}~\bibnamefont
  {Scarsoglio}}, \bibinfo {author} {\bibfnamefont {G.}~\bibnamefont
  {Iacobello}},\ and\ \bibinfo {author} {\bibfnamefont {L.}~\bibnamefont
  {Ridolfi}},\ }\bibfield  {title} {\enquote {\bibinfo {title} {Complex
  networks unveiling spatial patterns in turbulence},}\ }\href
  {https://doi.org/10.1142/S0218127416502230} {\bibfield  {journal} {\bibinfo
  {journal} {International Journal of Bifurcation and Chaos}\ }\textbf
  {\bibinfo {volume} {26}},\ \bibinfo {pages} {1650223} (\bibinfo {year}
  {2016})}\BibitemShut {NoStop}%
\bibitem [{\citenamefont {Schlueter-Kuck}\ and\ \citenamefont
  {Dabiri}(2017)}]{schlueter2017coherent}%
  \BibitemOpen
  \bibfield  {author} {\bibinfo {author} {\bibfnamefont {K.~L.}\ \bibnamefont
  {Schlueter-Kuck}}\ and\ \bibinfo {author} {\bibfnamefont {J.~O.}\
  \bibnamefont {Dabiri}},\ }\bibfield  {title} {\enquote {\bibinfo {title}
  {Coherent structure colouring: identification of coherent structures from
  sparse data using graph theory},}\ }\href
  {https://doi.org/10.1017/jfm.2016.755} {\bibfield  {journal} {\bibinfo
  {journal} {Journal of Fluid Mechanics}\ }\textbf {\bibinfo {volume} {811}},\
  \bibinfo {pages} {468--486} (\bibinfo {year} {2017})}\BibitemShut {NoStop}%
\bibitem [{\citenamefont {Vieweg}\ \emph {et~al.}(2021)\citenamefont {Vieweg},
  \citenamefont {Schneide}, \citenamefont {Padberg-Gehle},\ and\ \citenamefont
  {Schumacher}}]{vieweg}%
  \BibitemOpen
  \bibfield  {author} {\bibinfo {author} {\bibfnamefont {P.~P.}\ \bibnamefont
  {Vieweg}}, \bibinfo {author} {\bibfnamefont {C.}~\bibnamefont {Schneide}},
  \bibinfo {author} {\bibfnamefont {K.}~\bibnamefont {Padberg-Gehle}},\ and\
  \bibinfo {author} {\bibfnamefont {J.}~\bibnamefont {Schumacher}},\ }\bibfield
   {title} {\enquote {\bibinfo {title} {Lagrangian heat transport in turbulent
  three-dimensional convection},}\ }\href
  {https://doi.org/10.1103/PhysRevFluids.6.L041501} {\bibfield  {journal}
  {\bibinfo  {journal} {Phys. Rev. Fluids}\ }\textbf {\bibinfo {volume} {6}},\
  \bibinfo {pages} {L041501} (\bibinfo {year} {2021})}\BibitemShut {NoStop}%
\bibitem [{\citenamefont {Taira}, \citenamefont {Nair},\ and\ \citenamefont
  {Brunton}(2016)}]{taira2016jfm}%
  \BibitemOpen
  \bibfield  {author} {\bibinfo {author} {\bibfnamefont {K.}~\bibnamefont
  {Taira}}, \bibinfo {author} {\bibfnamefont {A.~G.}\ \bibnamefont {Nair}},\
  and\ \bibinfo {author} {\bibfnamefont {S.~L.}\ \bibnamefont {Brunton}},\
  }\bibfield  {title} {\enquote {\bibinfo {title} {Network structure of
  two-dimensional decaying isotropic turbulence},}\ }\href
  {https://doi.org/10.1017/jfm.2016.235} {\bibfield  {journal} {\bibinfo
  {journal} {Journal of Fluid Mechanics}\ }\textbf {\bibinfo {volume} {795}}
  (\bibinfo {year} {2016}),\ 10.1017/jfm.2016.235}\BibitemShut {NoStop}%
\bibitem [{\citenamefont {Yeh}, \citenamefont {Gopalakrishnan~Meena},\ and\
  \citenamefont {Taira}(2021)}]{yeh2021}%
  \BibitemOpen
  \bibfield  {author} {\bibinfo {author} {\bibfnamefont {C.-A.}\ \bibnamefont
  {Yeh}}, \bibinfo {author} {\bibfnamefont {M.}~\bibnamefont
  {Gopalakrishnan~Meena}},\ and\ \bibinfo {author} {\bibfnamefont
  {K.}~\bibnamefont {Taira}},\ }\bibfield  {title} {\enquote {\bibinfo {title}
  {Network broadcast analysis and control of turbulent flows},}\ }\href
  {https://doi.org/10.1017/jfm.2020.965} {\bibfield  {journal} {\bibinfo
  {journal} {Journal of Fluid Mechanics}\ }\textbf {\bibinfo {volume} {910}},\
  \bibinfo {pages} {A15} (\bibinfo {year} {2021})}\BibitemShut {NoStop}%
\bibitem [{\citenamefont {Rypina}\ and\ \citenamefont
  {Pratt}(2017)}]{rypina2017npg}%
  \BibitemOpen
  \bibfield  {author} {\bibinfo {author} {\bibfnamefont {I.~I.}\ \bibnamefont
  {Rypina}}\ and\ \bibinfo {author} {\bibfnamefont {L.~J.}\ \bibnamefont
  {Pratt}},\ }\bibfield  {title} {\enquote {\bibinfo {title} {Trajectory
  encounter volume as a diagnostic of mixing potential in fluid flows},}\
  }\href {https://doi.org/10.5194/npg-24-189-2017} {\bibfield  {journal}
  {\bibinfo  {journal} {Nonlinear Processes in Geophysics}\ }\textbf {\bibinfo
  {volume} {24}},\ \bibinfo {pages} {189--202} (\bibinfo {year}
  {2017})}\BibitemShut {NoStop}%
\bibitem [{\citenamefont {Banisch}, \citenamefont {Koltai},\ and\ \citenamefont
  {Padberg-Gehle}(2019)}]{banisch}%
  \BibitemOpen
  \bibfield  {author} {\bibinfo {author} {\bibfnamefont {R.}~\bibnamefont
  {Banisch}}, \bibinfo {author} {\bibfnamefont {P.}~\bibnamefont {Koltai}},\
  and\ \bibinfo {author} {\bibfnamefont {K.}~\bibnamefont {Padberg-Gehle}},\
  }\bibfield  {title} {\enquote {\bibinfo {title} {Network measures of
  mixing},}\ }\href {https://doi.org/10.1063/1.5087632} {\bibfield  {journal}
  {\bibinfo  {journal} {Chaos: An Interdisciplinary Journal of Nonlinear
  Science}\ }\textbf {\bibinfo {volume} {29}},\ \bibinfo {pages} {063125}
  (\bibinfo {year} {2019})}\BibitemShut {NoStop}%
\bibitem [{\citenamefont {Ser-Giacomi}\ \emph {et~al.}(2015)\citenamefont
  {Ser-Giacomi}, \citenamefont {Rossi}, \citenamefont {López},\ and\
  \citenamefont {Hernández-García}}]{sergiac15}%
  \BibitemOpen
  \bibfield  {author} {\bibinfo {author} {\bibfnamefont {E.}~\bibnamefont
  {Ser-Giacomi}}, \bibinfo {author} {\bibfnamefont {V.}~\bibnamefont {Rossi}},
  \bibinfo {author} {\bibfnamefont {C.}~\bibnamefont {López}},\ and\ \bibinfo
  {author} {\bibfnamefont {E.}~\bibnamefont {Hernández-García}},\ }\bibfield
  {title} {\enquote {\bibinfo {title} {Flow networks: A characterization of
  geophysical fluid transport},}\ }\href {https://doi.org/10.1063/1.4908231}
  {\bibfield  {journal} {\bibinfo  {journal} {Chaos: An Interdisciplinary
  Journal of Nonlinear Science}\ }\textbf {\bibinfo {volume} {25}},\ \bibinfo
  {pages} {036404} (\bibinfo {year} {2015})},\ \Eprint
  {https://arxiv.org/abs/https://doi.org/10.1063/1.4908231}
  {https://doi.org/10.1063/1.4908231} \BibitemShut {NoStop}%
\bibitem [{\citenamefont {Perrone}\ \emph {et~al.}(2020)\citenamefont
  {Perrone}, \citenamefont {Kuerten}, \citenamefont {Ridolfi},\ and\
  \citenamefont {Scarsoglio}}]{perrone20}%
  \BibitemOpen
  \bibfield  {author} {\bibinfo {author} {\bibfnamefont {D.}~\bibnamefont
  {Perrone}}, \bibinfo {author} {\bibfnamefont {J.~G.~M.}\ \bibnamefont
  {Kuerten}}, \bibinfo {author} {\bibfnamefont {L.}~\bibnamefont {Ridolfi}},\
  and\ \bibinfo {author} {\bibfnamefont {S.}~\bibnamefont {Scarsoglio}},\
  }\bibfield  {title} {\enquote {\bibinfo {title} {Wall-induced anisotropy
  effects on turbulent mixing in channel flow: A network-based analysis},}\
  }\href {https://doi.org/10.1103/PhysRevE.102.043109} {\bibfield  {journal}
  {\bibinfo  {journal} {Phys. Rev. E}\ }\textbf {\bibinfo {volume} {102}},\
  \bibinfo {pages} {043109} (\bibinfo {year} {2020})}\BibitemShut {NoStop}%
\bibitem [{\citenamefont {Perrone}\ \emph {et~al.}(2021)\citenamefont
  {Perrone}, \citenamefont {Kuerten}, \citenamefont {Ridolfi},\ and\
  \citenamefont {Scarsoglio}}]{perrone21}%
  \BibitemOpen
  \bibfield  {author} {\bibinfo {author} {\bibfnamefont {D.}~\bibnamefont
  {Perrone}}, \bibinfo {author} {\bibfnamefont {J.~G.~M.}\ \bibnamefont
  {Kuerten}}, \bibinfo {author} {\bibfnamefont {L.}~\bibnamefont {Ridolfi}},\
  and\ \bibinfo {author} {\bibfnamefont {S.}~\bibnamefont {Scarsoglio}},\
  }\bibfield  {title} {\enquote {\bibinfo {title} {Network analysis of reynolds
  number scaling in wall-bounded lagrangian mixing},}\ }\href
  {https://doi.org/10.1103/PhysRevFluids.6.124501} {\bibfield  {journal}
  {\bibinfo  {journal} {Phys. Rev. Fluids}\ }\textbf {\bibinfo {volume} {6}},\
  \bibinfo {pages} {124501} (\bibinfo {year} {2021})}\BibitemShut {NoStop}%
\bibitem [{\citenamefont {Donne}\ \emph {et~al.}(2011)\citenamefont {Donne},
  \citenamefont {Small}, \citenamefont {Donges}, \citenamefont {Marwan},
  \citenamefont {Zou}, \citenamefont {Xiang},\ and\ \citenamefont
  {Kurths}}]{donner11}%
  \BibitemOpen
  \bibfield  {author} {\bibinfo {author} {\bibfnamefont {R.~V.}\ \bibnamefont
  {Donne}}, \bibinfo {author} {\bibfnamefont {M.}~\bibnamefont {Small}},
  \bibinfo {author} {\bibfnamefont {J.~F.}\ \bibnamefont {Donges}}, \bibinfo
  {author} {\bibfnamefont {N.}~\bibnamefont {Marwan}}, \bibinfo {author}
  {\bibfnamefont {Y.}~\bibnamefont {Zou}}, \bibinfo {author} {\bibfnamefont
  {R.}~\bibnamefont {Xiang}},\ and\ \bibinfo {author} {\bibfnamefont
  {J.}~\bibnamefont {Kurths}},\ }\bibfield  {title} {\enquote {\bibinfo {title}
  {Recurrence-based time series analysis by means of complex network
  methods},}\ }\href {https://doi.org/10.1142/S0218127411029021} {\bibfield
  {journal} {\bibinfo  {journal} {International Journal of Bifurcation and
  Chaos}\ }\textbf {\bibinfo {volume} {21}},\ \bibinfo {pages} {1019--1046}
  (\bibinfo {year} {2011})},\ \Eprint
  {https://arxiv.org/abs/https://doi.org/10.1142/S0218127411029021}
  {https://doi.org/10.1142/S0218127411029021} \BibitemShut {NoStop}%
\bibitem [{\citenamefont {Godavarthi}\ \emph {et~al.}(2017)\citenamefont
  {Godavarthi}, \citenamefont {Unni}, \citenamefont {Gopalakrishnan},\ and\
  \citenamefont {Sujith}}]{godavarthi17}%
  \BibitemOpen
  \bibfield  {author} {\bibinfo {author} {\bibfnamefont {V.}~\bibnamefont
  {Godavarthi}}, \bibinfo {author} {\bibfnamefont {V.~R.}\ \bibnamefont
  {Unni}}, \bibinfo {author} {\bibfnamefont {E.~A.}\ \bibnamefont
  {Gopalakrishnan}},\ and\ \bibinfo {author} {\bibfnamefont {R.~I.}\
  \bibnamefont {Sujith}},\ }\bibfield  {title} {\enquote {\bibinfo {title}
  {Recurrence networks to study dynamical transitions in a turbulent
  combustor},}\ }\href {https://doi.org/10.1063/1.4985275} {\bibfield
  {journal} {\bibinfo  {journal} {Chaos: An Interdisciplinary Journal of
  Nonlinear Science}\ }\textbf {\bibinfo {volume} {27}},\ \bibinfo {pages}
  {063113} (\bibinfo {year} {2017})},\ \Eprint
  {https://arxiv.org/abs/https://doi.org/10.1063/1.4985275}
  {https://doi.org/10.1063/1.4985275} \BibitemShut {NoStop}%
\bibitem [{\citenamefont {Kasthuri}\ \emph {et~al.}(2019)\citenamefont
  {Kasthuri}, \citenamefont {Pavithran}, \citenamefont {Pawar}, \citenamefont
  {Sujith}, \citenamefont {Gejji},\ and\ \citenamefont
  {Anderson}}]{praveen2019}%
  \BibitemOpen
  \bibfield  {author} {\bibinfo {author} {\bibfnamefont {P.}~\bibnamefont
  {Kasthuri}}, \bibinfo {author} {\bibfnamefont {I.}~\bibnamefont {Pavithran}},
  \bibinfo {author} {\bibfnamefont {S.~A.}\ \bibnamefont {Pawar}}, \bibinfo
  {author} {\bibfnamefont {R.~I.}\ \bibnamefont {Sujith}}, \bibinfo {author}
  {\bibfnamefont {R.}~\bibnamefont {Gejji}},\ and\ \bibinfo {author}
  {\bibfnamefont {W.}~\bibnamefont {Anderson}},\ }\bibfield  {title} {\enquote
  {\bibinfo {title} {Dynamical systems approach to study thermoacoustic
  transitions in a liquid rocket combustor},}\ }\href
  {https://doi.org/10.1063/1.5120429} {\bibfield  {journal} {\bibinfo
  {journal} {Chaos: An Interdisciplinary Journal of Nonlinear Science}\
  }\textbf {\bibinfo {volume} {29}},\ \bibinfo {pages} {103115} (\bibinfo
  {year} {2019})},\ \Eprint
  {https://arxiv.org/abs/https://doi.org/10.1063/1.5120429}
  {https://doi.org/10.1063/1.5120429} \BibitemShut {NoStop}%
\bibitem [{\citenamefont {Tandon}\ and\ \citenamefont
  {Sujith}(2021)}]{tandon21}%
  \BibitemOpen
  \bibfield  {author} {\bibinfo {author} {\bibfnamefont {S.}~\bibnamefont
  {Tandon}}\ and\ \bibinfo {author} {\bibfnamefont {R.~I.}\ \bibnamefont
  {Sujith}},\ }\bibfield  {title} {\enquote {\bibinfo {title} {Condensation in
  the phase space and network topology during transition from chaos to order in
  turbulent thermoacoustic systems},}\ }\href
  {https://doi.org/10.1063/5.0039229} {\bibfield  {journal} {\bibinfo
  {journal} {Chaos: An Interdisciplinary Journal of Nonlinear Science}\
  }\textbf {\bibinfo {volume} {31}},\ \bibinfo {pages} {043126} (\bibinfo
  {year} {2021})},\ \Eprint
  {https://arxiv.org/abs/https://doi.org/10.1063/5.0039229}
  {https://doi.org/10.1063/5.0039229} \BibitemShut {NoStop}%
\bibitem [{\citenamefont {Nomi}\ \emph {et~al.}(2021)\citenamefont {Nomi},
  \citenamefont {Gotoda}, \citenamefont {Fukuda},\ and\ \citenamefont
  {Almarcha}}]{gotoda2}%
  \BibitemOpen
  \bibfield  {author} {\bibinfo {author} {\bibfnamefont {Y.}~\bibnamefont
  {Nomi}}, \bibinfo {author} {\bibfnamefont {H.}~\bibnamefont {Gotoda}},
  \bibinfo {author} {\bibfnamefont {S.}~\bibnamefont {Fukuda}},\ and\ \bibinfo
  {author} {\bibfnamefont {C.}~\bibnamefont {Almarcha}},\ }\bibfield  {title}
  {\enquote {\bibinfo {title} {Complex network analysis of spatiotemporal
  dynamics of premixed flame in a hele-shaw cell: A transition from chaos to
  stochastic state},}\ }\href {https://doi.org/10.1063/5.0070526} {\bibfield
  {journal} {\bibinfo  {journal} {Chaos: An Interdisciplinary Journal of
  Nonlinear Science}\ }\textbf {\bibinfo {volume} {31}},\ \bibinfo {pages}
  {123133} (\bibinfo {year} {2021})},\ \Eprint
  {https://arxiv.org/abs/https://doi.org/10.1063/5.0070526}
  {https://doi.org/10.1063/5.0070526} \BibitemShut {NoStop}%
\bibitem [{\citenamefont {Hachijo}\ \emph {et~al.}(2020)\citenamefont
  {Hachijo}, \citenamefont {Gotoda}, \citenamefont {Nishizawa},\ and\
  \citenamefont {Kazawa}}]{gotoda3}%
  \BibitemOpen
  \bibfield  {author} {\bibinfo {author} {\bibfnamefont {T.}~\bibnamefont
  {Hachijo}}, \bibinfo {author} {\bibfnamefont {H.}~\bibnamefont {Gotoda}},
  \bibinfo {author} {\bibfnamefont {T.}~\bibnamefont {Nishizawa}},\ and\
  \bibinfo {author} {\bibfnamefont {J.}~\bibnamefont {Kazawa}},\ }\bibfield
  {title} {\enquote {\bibinfo {title} {Early detection of cascade flutter in a
  model aircraft turbine using a methodology combining complex networks and
  synchronization},}\ }\href {https://doi.org/10.1103/PhysRevApplied.14.014093}
  {\bibfield  {journal} {\bibinfo  {journal} {Phys. Rev. Applied}\ }\textbf
  {\bibinfo {volume} {14}},\ \bibinfo {pages} {014093} (\bibinfo {year}
  {2020})}\BibitemShut {NoStop}%
\bibitem [{\citenamefont {Lacasa}\ \emph {et~al.}(2008)\citenamefont {Lacasa},
  \citenamefont {Luque}, \citenamefont {Ballesteros}, \citenamefont {Luque},\
  and\ \citenamefont {Nuno}}]{lacasa2008}%
  \BibitemOpen
  \bibfield  {author} {\bibinfo {author} {\bibfnamefont {L.}~\bibnamefont
  {Lacasa}}, \bibinfo {author} {\bibfnamefont {B.}~\bibnamefont {Luque}},
  \bibinfo {author} {\bibfnamefont {F.}~\bibnamefont {Ballesteros}}, \bibinfo
  {author} {\bibfnamefont {J.}~\bibnamefont {Luque}},\ and\ \bibinfo {author}
  {\bibfnamefont {J.~C.}\ \bibnamefont {Nuno}},\ }\bibfield  {title} {\enquote
  {\bibinfo {title} {From time series to complex networks: The visibility
  graph},}\ }\href {https://doi.org/10.1073/pnas.0709247105} {\bibfield
  {journal} {\bibinfo  {journal} {Proceedings of the National Academy of
  Sciences}\ }\textbf {\bibinfo {volume} {105}},\ \bibinfo {pages} {4972--4975}
  (\bibinfo {year} {2008})},\ \Eprint
  {https://arxiv.org/abs/https://www.pnas.org/doi/pdf/10.1073/pnas.0709247105}
  {https://www.pnas.org/doi/pdf/10.1073/pnas.0709247105} \BibitemShut {NoStop}%
\bibitem [{\citenamefont {Iacobello}, \citenamefont {Scarsoglio},\ and\
  \citenamefont {Ridolfi}(2018)}]{iacobello2018}%
  \BibitemOpen
  \bibfield  {author} {\bibinfo {author} {\bibfnamefont {G.}~\bibnamefont
  {Iacobello}}, \bibinfo {author} {\bibfnamefont {S.}~\bibnamefont
  {Scarsoglio}},\ and\ \bibinfo {author} {\bibfnamefont {L.}~\bibnamefont
  {Ridolfi}},\ }\bibfield  {title} {\enquote {\bibinfo {title} {Visibility
  graph analysis of wall turbulence time-series},}\ }\href
  {https://doi.org/https://doi.org/10.1016/j.physleta.2017.10.027} {\bibfield
  {journal} {\bibinfo  {journal} {Physics Letters A}\ }\textbf {\bibinfo
  {volume} {382}},\ \bibinfo {pages} {1--11} (\bibinfo {year}
  {2018})}\BibitemShut {NoStop}%
\bibitem [{\citenamefont {Iacobello}\ \emph {et~al.}(2019)\citenamefont
  {Iacobello}, \citenamefont {Marro}, \citenamefont {Ridolfi}, \citenamefont
  {Salizzoni},\ and\ \citenamefont {Scarsoglio}}]{iacobello2019}%
  \BibitemOpen
  \bibfield  {author} {\bibinfo {author} {\bibfnamefont {G.}~\bibnamefont
  {Iacobello}}, \bibinfo {author} {\bibfnamefont {M.}~\bibnamefont {Marro}},
  \bibinfo {author} {\bibfnamefont {L.}~\bibnamefont {Ridolfi}}, \bibinfo
  {author} {\bibfnamefont {P.}~\bibnamefont {Salizzoni}},\ and\ \bibinfo
  {author} {\bibfnamefont {S.}~\bibnamefont {Scarsoglio}},\ }\bibfield  {title}
  {\enquote {\bibinfo {title} {Experimental investigation of vertical turbulent
  transport of a passive scalar in a boundary layer: Statistics and visibility
  graph analysis},}\ }\href {https://doi.org/10.1103/PhysRevFluids.4.104501}
  {\bibfield  {journal} {\bibinfo  {journal} {Phys. Rev. Fluids}\ }\textbf
  {\bibinfo {volume} {4}},\ \bibinfo {pages} {104501} (\bibinfo {year}
  {2019})}\BibitemShut {NoStop}%
\bibitem [{\citenamefont {Chowdhuri}, \citenamefont {Iacobello},\ and\
  \citenamefont {Banerjee}(2021)}]{chowdhuri2021}%
  \BibitemOpen
  \bibfield  {author} {\bibinfo {author} {\bibfnamefont {S.}~\bibnamefont
  {Chowdhuri}}, \bibinfo {author} {\bibfnamefont {G.}~\bibnamefont
  {Iacobello}},\ and\ \bibinfo {author} {\bibfnamefont {T.}~\bibnamefont
  {Banerjee}},\ }\bibfield  {title} {\enquote {\bibinfo {title} {Visibility
  network analysis of large-scale intermittency in convective surface layer
  turbulence},}\ }\href {https://doi.org/10.1017/jfm.2021.720} {\bibfield
  {journal} {\bibinfo  {journal} {Journal of Fluid Mechanics}\ }\textbf
  {\bibinfo {volume} {925}},\ \bibinfo {pages} {A38} (\bibinfo {year}
  {2021})}\BibitemShut {NoStop}%
\bibitem [{\citenamefont {Iacobello}, \citenamefont {Ridolfi},\ and\
  \citenamefont {Scarsoglio}(2021{\natexlab{b}})}]{iacobello2021a}%
  \BibitemOpen
  \bibfield  {author} {\bibinfo {author} {\bibfnamefont {G.}~\bibnamefont
  {Iacobello}}, \bibinfo {author} {\bibfnamefont {L.}~\bibnamefont {Ridolfi}},\
  and\ \bibinfo {author} {\bibfnamefont {S.}~\bibnamefont {Scarsoglio}},\
  }\bibfield  {title} {\enquote {\bibinfo {title} {Large-to-small scale
  frequency modulation analysis in wall-bounded turbulence via visibility
  networks},}\ }\href {https://doi.org/10.1017/jfm.2021.279} {\bibfield
  {journal} {\bibinfo  {journal} {Journal of Fluid Mechanics}\ }\textbf
  {\bibinfo {volume} {918}},\ \bibinfo {pages} {A13} (\bibinfo {year}
  {2021}{\natexlab{b}})}\BibitemShut {NoStop}%
\bibitem [{\citenamefont {Kobayashi}\ \emph {et~al.}(2019)\citenamefont
  {Kobayashi}, \citenamefont {Gotoda}, \citenamefont {Kandani}, \citenamefont
  {Ohmichi},\ and\ \citenamefont {Matsuyama}}]{gotoda1}%
  \BibitemOpen
  \bibfield  {author} {\bibinfo {author} {\bibfnamefont {W.}~\bibnamefont
  {Kobayashi}}, \bibinfo {author} {\bibfnamefont {H.}~\bibnamefont {Gotoda}},
  \bibinfo {author} {\bibfnamefont {S.}~\bibnamefont {Kandani}}, \bibinfo
  {author} {\bibfnamefont {Y.}~\bibnamefont {Ohmichi}},\ and\ \bibinfo {author}
  {\bibfnamefont {S.}~\bibnamefont {Matsuyama}},\ }\bibfield  {title} {\enquote
  {\bibinfo {title} {Spatiotemporal dynamics of turbulent coaxial jet analyzed
  by symbolic information-theory quantifiers and complex-network approach},}\
  }\href {https://doi.org/10.1063/1.5126490} {\bibfield  {journal} {\bibinfo
  {journal} {Chaos: An Interdisciplinary Journal of Nonlinear Science}\
  }\textbf {\bibinfo {volume} {29}},\ \bibinfo {pages} {123110} (\bibinfo
  {year} {2019})},\ \Eprint
  {https://arxiv.org/abs/https://doi.org/10.1063/1.5126490}
  {https://doi.org/10.1063/1.5126490} \BibitemShut {NoStop}%
\bibitem [{\citenamefont {Li}\ \emph {et~al.}(2008)\citenamefont {Li},
  \citenamefont {Perlman}, \citenamefont {Wan}, \citenamefont {Yang},
  \citenamefont {Meneveau}, \citenamefont {Burns}, \citenamefont {Chen},
  \citenamefont {Szalay},\ and\ \citenamefont {Eyink}}]{jhtdb1}%
  \BibitemOpen
  \bibfield  {author} {\bibinfo {author} {\bibfnamefont {Y.}~\bibnamefont
  {Li}}, \bibinfo {author} {\bibfnamefont {E.}~\bibnamefont {Perlman}},
  \bibinfo {author} {\bibfnamefont {M.}~\bibnamefont {Wan}}, \bibinfo {author}
  {\bibfnamefont {Y.}~\bibnamefont {Yang}}, \bibinfo {author} {\bibfnamefont
  {C.}~\bibnamefont {Meneveau}}, \bibinfo {author} {\bibfnamefont
  {R.}~\bibnamefont {Burns}}, \bibinfo {author} {\bibfnamefont
  {S.}~\bibnamefont {Chen}}, \bibinfo {author} {\bibfnamefont {A.}~\bibnamefont
  {Szalay}},\ and\ \bibinfo {author} {\bibfnamefont {G.}~\bibnamefont
  {Eyink}},\ }\bibfield  {title} {\enquote {\bibinfo {title} {A public
  turbulence database cluster and applications to study lagrangian evolution of
  velocity increments in turbulence},}\ }\href
  {https://doi.org/10.1080/14685240802376389} {\bibfield  {journal} {\bibinfo
  {journal} {Journal of Turbulence}\ }\textbf {\bibinfo {volume} {9}},\
  \bibinfo {pages} {N31} (\bibinfo {year} {2008})},\ \Eprint
  {https://arxiv.org/abs/https://doi.org/10.1080/14685240802376389}
  {https://doi.org/10.1080/14685240802376389} \BibitemShut {NoStop}%
\bibitem [{\citenamefont {Perlman}\ \emph {et~al.}(2007)\citenamefont
  {Perlman}, \citenamefont {Burns}, \citenamefont {Li},\ and\ \citenamefont
  {Meneveau}}]{jhtdb2}%
  \BibitemOpen
  \bibfield  {author} {\bibinfo {author} {\bibfnamefont {E.}~\bibnamefont
  {Perlman}}, \bibinfo {author} {\bibfnamefont {R.}~\bibnamefont {Burns}},
  \bibinfo {author} {\bibfnamefont {Y.}~\bibnamefont {Li}},\ and\ \bibinfo
  {author} {\bibfnamefont {C.}~\bibnamefont {Meneveau}},\ }\bibfield  {title}
  {\enquote {\bibinfo {title} {Data exploration of turbulence simulations using
  a database cluster},}\ }in\ \href {https://doi.org/10.1145/1362622.1362654}
  {\emph {\bibinfo {booktitle} {SC '07: Proceedings of the 2007 ACM/IEEE
  Conference on Supercomputing}}}\ (\bibinfo {year} {2007})\ pp.\ \bibinfo
  {pages} {1--11}\BibitemShut {NoStop}%
\bibitem [{\citenamefont {Watts}\ and\ \citenamefont
  {Strogatz}(1998)}]{watts1998}%
  \BibitemOpen
  \bibfield  {author} {\bibinfo {author} {\bibfnamefont {D.~J.}\ \bibnamefont
  {Watts}}\ and\ \bibinfo {author} {\bibfnamefont {S.~H.}\ \bibnamefont
  {Strogatz}},\ }\bibfield  {title} {\enquote {\bibinfo {title} {Collective
  dynamics of ‘small-world’networks},}\ }\href
  {https://doi.org/https://doi.org/10.1038/30918} {\bibfield  {journal}
  {\bibinfo  {journal} {nature}\ }\textbf {\bibinfo {volume} {393}},\ \bibinfo
  {pages} {440--442} (\bibinfo {year} {1998})}\BibitemShut {NoStop}%
\bibitem [{\citenamefont {Newman}(2002)}]{newman2002}%
  \BibitemOpen
  \bibfield  {author} {\bibinfo {author} {\bibfnamefont {M.~E.~J.}\
  \bibnamefont {Newman}},\ }\bibfield  {title} {\enquote {\bibinfo {title}
  {Assortative mixing in networks},}\ }\href
  {https://doi.org/10.1103/PhysRevLett.89.208701} {\bibfield  {journal}
  {\bibinfo  {journal} {Phys. Rev. Lett.}\ }\textbf {\bibinfo {volume} {89}},\
  \bibinfo {pages} {208701} (\bibinfo {year} {2002})}\BibitemShut {NoStop}%
\bibitem [{\citenamefont {Newman}(2003)}]{newman2003}%
  \BibitemOpen
  \bibfield  {author} {\bibinfo {author} {\bibfnamefont {M.~E.~J.}\
  \bibnamefont {Newman}},\ }\bibfield  {title} {\enquote {\bibinfo {title}
  {Mixing patterns in networks},}\ }\href
  {https://doi.org/10.1103/PhysRevE.67.026126} {\bibfield  {journal} {\bibinfo
  {journal} {Phys. Rev. E}\ }\textbf {\bibinfo {volume} {67}},\ \bibinfo
  {pages} {026126} (\bibinfo {year} {2003})}\BibitemShut {NoStop}%
\bibitem [{\citenamefont {Murugesan}\ and\ \citenamefont
  {Sujith}(2015)}]{murugesan2015}%
  \BibitemOpen
  \bibfield  {author} {\bibinfo {author} {\bibfnamefont {M.}~\bibnamefont
  {Murugesan}}\ and\ \bibinfo {author} {\bibfnamefont {R.}~\bibnamefont
  {Sujith}},\ }\bibfield  {title} {\enquote {\bibinfo {title} {Combustion noise
  is scale-free: transition from scale-free to order at the onset of
  thermoacoustic instability},}\ }\href {https://doi.org/10.1017/jfm.2015.215}
  {\bibfield  {journal} {\bibinfo  {journal} {Journal of Fluid Mechanics}\
  }\textbf {\bibinfo {volume} {772}},\ \bibinfo {pages} {225--245} (\bibinfo
  {year} {2015})}\BibitemShut {NoStop}%
\bibitem [{\citenamefont {Singh}\ \emph {et~al.}(2017)\citenamefont {Singh},
  \citenamefont {Belur~Vishwanath}, \citenamefont {Chaudhuri},\ and\
  \citenamefont {Sujith}}]{singh2017}%
  \BibitemOpen
  \bibfield  {author} {\bibinfo {author} {\bibfnamefont {J.}~\bibnamefont
  {Singh}}, \bibinfo {author} {\bibfnamefont {R.}~\bibnamefont
  {Belur~Vishwanath}}, \bibinfo {author} {\bibfnamefont {S.}~\bibnamefont
  {Chaudhuri}},\ and\ \bibinfo {author} {\bibfnamefont {R.~I.}\ \bibnamefont
  {Sujith}},\ }\bibfield  {title} {\enquote {\bibinfo {title} {Network
  structure of turbulent premixed flames},}\ }\href
  {https://doi.org/10.1063/1.4980135} {\bibfield  {journal} {\bibinfo
  {journal} {Chaos: An Interdisciplinary Journal of Nonlinear Science}\
  }\textbf {\bibinfo {volume} {27}},\ \bibinfo {pages} {043107} (\bibinfo
  {year} {2017})},\ \Eprint
  {https://arxiv.org/abs/https://doi.org/10.1063/1.4980135}
  {https://doi.org/10.1063/1.4980135} \BibitemShut {NoStop}%
\bibitem [{\citenamefont {Lan}\ \emph {et~al.}(2015)\citenamefont {Lan},
  \citenamefont {Mo}, \citenamefont {Chen}, \citenamefont {Liu},\ and\
  \citenamefont {Deng}}]{lan2015}%
  \BibitemOpen
  \bibfield  {author} {\bibinfo {author} {\bibfnamefont {X.}~\bibnamefont
  {Lan}}, \bibinfo {author} {\bibfnamefont {H.}~\bibnamefont {Mo}}, \bibinfo
  {author} {\bibfnamefont {S.}~\bibnamefont {Chen}}, \bibinfo {author}
  {\bibfnamefont {Q.}~\bibnamefont {Liu}},\ and\ \bibinfo {author}
  {\bibfnamefont {Y.}~\bibnamefont {Deng}},\ }\bibfield  {title} {\enquote
  {\bibinfo {title} {Fast transformation from time series to visibility
  graphs},}\ }\href {https://doi.org/10.1063/1.4927835} {\bibfield  {journal}
  {\bibinfo  {journal} {Chaos: An Interdisciplinary Journal of Nonlinear
  Science}\ }\textbf {\bibinfo {volume} {25}},\ \bibinfo {pages} {083105}
  (\bibinfo {year} {2015})},\ \Eprint
  {https://arxiv.org/abs/https://doi.org/10.1063/1.4927835}
  {https://doi.org/10.1063/1.4927835} \BibitemShut {NoStop}%
\bibitem [{\citenamefont {Martin}\ \emph {et~al.}(2011)\citenamefont {Martin},
  \citenamefont {Brown}, \citenamefont {Klavans},\ and\ \citenamefont
  {Boyack}}]{openord}%
  \BibitemOpen
  \bibfield  {author} {\bibinfo {author} {\bibfnamefont {S.}~\bibnamefont
  {Martin}}, \bibinfo {author} {\bibfnamefont {W.~M.}\ \bibnamefont {Brown}},
  \bibinfo {author} {\bibfnamefont {R.}~\bibnamefont {Klavans}},\ and\ \bibinfo
  {author} {\bibfnamefont {K.~W.}\ \bibnamefont {Boyack}},\ }\bibfield  {title}
  {\enquote {\bibinfo {title} {{OpenOrd: an open-source toolbox for large graph
  layout}},}\ }in\ \href {https://doi.org/10.1117/12.871402} {\emph {\bibinfo
  {booktitle} {Visualization and Data Analysis 2011}}},\ Vol.\ \bibinfo
  {volume} {7868},\ \bibinfo {editor} {edited by\ \bibinfo {editor}
  {\bibfnamefont {P.~C.}\ \bibnamefont {Wong}}, \bibinfo {editor}
  {\bibfnamefont {J.}~\bibnamefont {Park}}, \bibinfo {editor} {\bibfnamefont
  {M.~C.}\ \bibnamefont {Hao}}, \bibinfo {editor} {\bibfnamefont
  {C.}~\bibnamefont {Chen}}, \bibinfo {editor} {\bibfnamefont {K.}~\bibnamefont
  {Börner}}, \bibinfo {editor} {\bibfnamefont {D.~L.}\ \bibnamefont {Kao}},\
  and\ \bibinfo {editor} {\bibfnamefont {J.~C.}\ \bibnamefont {Roberts}}},\
  \bibinfo {organization} {International Society for Optics and Photonics}\
  (\bibinfo  {publisher} {SPIE},\ \bibinfo {year} {2011})\ pp.\ \bibinfo
  {pages} {45 -- 55}\BibitemShut {NoStop}%
\bibitem [{\citenamefont {Blondel}\ \emph {et~al.}(2008)\citenamefont
  {Blondel}, \citenamefont {Guillaume}, \citenamefont {Lambiotte},\ and\
  \citenamefont {Lefebvre}}]{louvain}%
  \BibitemOpen
  \bibfield  {author} {\bibinfo {author} {\bibfnamefont {V.~D.}\ \bibnamefont
  {Blondel}}, \bibinfo {author} {\bibfnamefont {J.-L.}\ \bibnamefont
  {Guillaume}}, \bibinfo {author} {\bibfnamefont {R.}~\bibnamefont
  {Lambiotte}},\ and\ \bibinfo {author} {\bibfnamefont {E.}~\bibnamefont
  {Lefebvre}},\ }\bibfield  {title} {\enquote {\bibinfo {title} {Fast unfolding
  of communities in large networks},}\ }\href
  {https://doi.org/10.1088/1742-5468/2008/10/p10008} {\bibfield  {journal}
  {\bibinfo  {journal} {Journal of Statistical Mechanics: Theory and
  Experiment}\ }\textbf {\bibinfo {volume} {2008}},\ \bibinfo {pages} {P10008}
  (\bibinfo {year} {2008})}\BibitemShut {NoStop}%
\bibitem [{\citenamefont {Lacasa}\ \emph {et~al.}(2012)\citenamefont {Lacasa},
  \citenamefont {Nunez}, \citenamefont {Rold{\'a}n}, \citenamefont {Parrondo},\
  and\ \citenamefont {Luque}}]{lacasa2012time}%
  \BibitemOpen
  \bibfield  {author} {\bibinfo {author} {\bibfnamefont {L.}~\bibnamefont
  {Lacasa}}, \bibinfo {author} {\bibfnamefont {A.}~\bibnamefont {Nunez}},
  \bibinfo {author} {\bibfnamefont {{\'E}.}~\bibnamefont {Rold{\'a}n}},
  \bibinfo {author} {\bibfnamefont {J.~M.}\ \bibnamefont {Parrondo}},\ and\
  \bibinfo {author} {\bibfnamefont {B.}~\bibnamefont {Luque}},\ }\bibfield
  {title} {\enquote {\bibinfo {title} {Time series irreversibility: a
  visibility graph approach},}\ }\href
  {https://doi.org/10.1140/epjb/e2012-20809-8} {\bibfield  {journal} {\bibinfo
  {journal} {The European Physical Journal B}\ }\textbf {\bibinfo {volume}
  {85}},\ \bibinfo {pages} {1--11} (\bibinfo {year} {2012})}\BibitemShut
  {NoStop}%
\end{thebibliography}%

\end{document}